\documentclass[journal=jacsat,manuscript=article]{achemso}

\usepackage{achemso}
\usepackage{xcolor}
\usepackage{amsmath}
\usepackage{threeparttable}
\usepackage[table]{xcolor}
\usepackage{placeins}
\usepackage{amsmath,amssymb,amsfonts}
\usepackage{algorithmic}
\usepackage{graphicx}
\usepackage{textcomp}
\usepackage{xcolor}
\usepackage{textgreek}
\usepackage{booktabs}
\usepackage{multirow}
\usepackage{caption}
\usepackage{subcaption}
\usepackage{siunitx}
\usepackage{mathptmx}
\usepackage{gensymb}

\usepackage{lineno}

\usepackage{xr}

\usepackage{mciteplus}

\floatstyle{plain}
\restylefloat{figure}

\author{Guillermo L. Esparza}

\affiliation[UCSDNANO]
{Aiiso Yufeng Li Family Department of Chemical and Nano Engineering,
University of California,
San Diego, La Jolla, CA 92093, USA}

\author{Zhenkun Yuan}
\affiliation[DC]
{Thayer School of Engineering, Dartmouth College, Hanover, NH 03755, USA}

\author{Muhammad Rubaiat Hasan}

\affiliation[ISU]
{Department of Chemistry, Iowa State University, Ames, IA 50011, USA}

\author{Yagmur Coban}

\affiliation[DC]
{Thayer School of Engineering, Dartmouth College, Hanover, NH 03755, USA}

\author{Gideon Kassa}

\affiliation[DC]
{Thayer School of Engineering, Dartmouth College, Hanover, NH 03755, USA}

\author{Vivek Shastry Devalla}

\affiliation[UCSDNANO]
{Aiiso Yufeng Li Family Department of Chemical and Nano Engineering, University of California, San Diego, La Jolla, CA 92093, USA}
\alsoaffiliation[UCI]
{Department of Chemical and Biomolecular Engineering, University of California, Irvine, CA 92697, USA }

\author{Tejas Nivarty}

\affiliation[UCSDNANO]
{Aiiso Yufeng Li Family Department of Chemical and Nano Engineering, University of California, San Diego, La Jolla, CA 92093, USA}

\author{Jack R. Palmer}

\affiliation[UCSDNANO]
{Aiiso Yufeng Li Family Department of Chemical and Nano Engineering, University of California, San Diego, La Jolla, CA 92093, USA}

\author{Jifeng Liu}

\affiliation[DC]
{Thayer School of Engineering, Dartmouth College, Hanover, NH 03755, USA}

\author{Kirill Kovnir}

\affiliation[ISU]
{Department of Chemistry, Iowa State University, Ames, IA 50011, USA}

\alsoaffiliation[ANL]
{Ames National Laboratory, U.S. Department of Energy, Ames, IA 50011, USA}

\author{Geoffroy Hautier}

\affiliation[DC]
{Thayer School of Engineering, Dartmouth College, Hanover, NH 03755, USA}

\alsoaffiliation[RU]
{Materials Science and NanoEngineering, Rice University, Houston, TX 77005, USA}

\author{David P. Fenning}

\email{dfenning@ucsd.edu}

\affiliation[UCSDNANO]
{Aiiso Yufeng Li Family Department of Chemical and Nano Engineering, University of California, San Diego, La Jolla, CA 92093, USA}

\alsoaffiliation[UCSDMATS]
{Materials Science \& Engineering Program, University of California, San Diego, La Jolla, CA 92093, USA}

\title[CCP title]
 {CaCd$_2$P$_2$: A Visible-Light Absorbing Zintl Phosphide Stable under Photoelectrochemical Water Oxidation}

\begin{document}

\newpage{}

\begin{abstract}

A key bottleneck to solar fuels is the absence of stable and strongly absorbing photoelectrode materials for the oxygen evolution reaction (OER). Modern approaches generally trade off between stable but weakly absorbing materials, such as wide bandgap oxides, or strongly absorbing materials that rely on encapsulation for stability and are weakly catalytic, such as the III-V family of semiconductors. Of interest are materials like transition metal phosphides, such as FeP$_2$, that are known to undergo beneficial in situ surface transformations in the oxidative environment of OER, though stability has remained a primary hurdle. Here we report on CaCd$_2$P$_2$, a Zintl phase visible-light absorber with favorable 1.6 eV bandgap, that we identified using high-throughput computational screening. Using a combination of photoelectrochemical measurements, microscopy, and spectroscopy, we show that CaCd$_2$P$_2$ undergoes a light-stabilized surface transformation that renders it stable under alkaline OER conditions. We also show that the well known OER catalyst CoPi can act as a stable co-catalyst in synergy with the \textit{in-situ} CaCd$_2$P$_2$ surface. The light-induced stabilization that CaCd$_2$P$_2$ displays is in sharp contrast to the photocorrosion commonly observed in visible light-absorbing photoelectrodes. The broader AM$_2$P$_2$ family of Zintl phases offers a significant opportunity to explore stabilizing interface chemistry and re-design the manner in which low-bandgap semiconductors are used for photoelectrochemical energy conversion.

\end{abstract}

\newpage

The seminal work by Fujishima and Honda using TiO$_2$ to leverage solar power for the direct generation of hydrogen \cite{Fujishima1972} launched an ongoing and decades long search for materials that meet the confluence of physical and chemical properties required for efficient solar fuel production. The requirement for strong absorption of photons, the generation, separation and transport of electrons and holes, and the catalytic assembly of fuel products at active sites at a single solid-electrolyte interface often create countervailing demands. In addition, photoelectrochemical (PEC) materials often experience conditions particularly conducive to photocorrosion given the presence of both electrons and holes near the electrolyte interface. In the context of the oxygen evolution reaction (OER), the 4-electron process involves multiple proton-coupled electron transfer steps, sluggish kinetics, and high over-potentials, all of which often lead to unwanted photocorrosion.

To address these challenges, researchers have explored a variety of different strategies. Foremost, the use of wide-bandgap materials, such as TiO$_2$ ($E_g > 3$ eV) and other oxides, can lead to photoelectrodes with high stability in OER conditions. However, this stability comes at the expense of effective use of the solar spectrum, where the wide bandgap results in the vast majority of solar photons going unabsorbed and therefore unable to do chemical work\cite{Schichtl2025, Lin2021}. Recent developments in the field focus on using secondary sources of illumination (such as blue light-emitting-diodes) to enhance the photocatalytic effect of these absorbers, but such approaches have the downside of more energy-conversion steps and increased complexity\cite{Schroeder2022}. 

The natural strategy to improve solar absorption is to use an absorber with a narrower bandgap. To this end, Si ($E_g =1.12$ eV) and III-V semiconductors (GaAs, $E_g = 1.42$ eV
) have been extensively studied for PEC applications\cite{Vilanova2024, Varadhan2019, Hallstrom2021, Schichtl2025, Bae2024}. However, these materials have low intrinsic catalytic activity and, to this day, no low-bandgap semiconductor has shown robust stability in OER when the unaltered semiconductor makes direct contact with the electrolyte. Rather, researchers have generally relied on careful encapsulation, where thin films of one-or-more materials (typically including a co-catalyst) completely protect the semiconductor from the electrolyte\cite{Hu2014, Choi2020a, Cao2020, Sivagurunathan2021a, Arunachalam2023}. Unfortunately, such strategies are prone to pin-hole defects, particularly when considering larger format devices, and rely on deposition techniques (such as atomic-layer deposition) that have proven challenging to scale within the low-cost, high-throughput manufacturing paradigm required for energy production.

Among low-bandgap semiconductors of interest are certain binary transition metal phosphides. In addition to their relative earth-abundance, the attention transition metal phosphides have garnered is driven in large part by the interesting phenomena that occur at their interfaces\cite{Serov2021,Aziz2023, Schichtl2025,Yang2024}. Specifically, they often undergo \emph{in situ} surface transformations, leading to new phases (i.e. oxides, oxy/hydroxides, and phosphates) that display enhanced activity and/or stability\cite{Dutta2017, Li2016, Li2019}. While the body of research into the self-passivation and pre-catalyst functionality of transition metal phosphides is now quite rich, the stability of these materials remains the main challenge for the field\cite{Aziz2023}.

Driven by high-throughput computational efforts for materials discovery, here we report the discovery of a 
Zintl phosphide, CaCd$_2$P$_2$, as a visible-light absorber ($E_g \approx 1.6 \text{ eV}$) with exceptional photostability under OER conditions in alkaline media (pH 13, 0.1 M KOH, simulated 1-sun conditions). We show that, rather than photocorroding, this material undergoes a light-stabilized surface transformation that robustly protects the semiconductor while allowing for charge transfer and OER. 
The OER performance of the neat semiconductor is modest, with a photovoltage and current up to $109\pm3$ mV and $119\pm 3$ $\mu$A/cm$^2_\text{geo}$ at 1.7 V$_\text{RHE}$ in our observations. 
We demonstrate its synergistic effects on both activity and stability with the well known Co-based co-catalyst, CoPi\cite{Kanan2008, Kanan2009, McCrory2013, Vensaus2024}. By electrodepositing a small amount of CoPi on CaCd$_2$P$_2$, we achieve a 1-2 order of magnitude improvement in catalytic activity and reduced overpotentials. We also show that the CaCd$_2$P$_2$ and CoPi pairing display significantly enhanced co-stability in alkaline media, particularly under illumination where a 100\% Faradaic efficiency is achieved for OER within error after activation. 
The identification of CaCd$_2$P$_2$ as a stable photoelectrode material in PEC OER motivates the exploration of the photoactive ternary Zintl phosphides and their alloys for solar fuel production.

\section{Computation, Synthesis, and Photophysics of CaCd$_2$P$_2$}

Recently, a large-scale computational materials screening for solar energy conversion applications highlighted a new class of light-absorbing semiconductors, the AM$_2$P$_2$ (A = Ca, Sr, Ba, M = Zn, Cd, Mg) ternary Zintl phosphides\cite{Yuan2024, Pike2025}. More specifically, this class of materials combine high visible-light absorption, high carrier mobility, and long carrier lifetime 
due to the absence of low-energy, deep intrinsic defects.
Subsequently, at least six of these materials have been synthesized in powder or thin-film formats as well as colloidal quantum dots, showing strong band-to-band photoluminescence (PL) emissions and measured long carrier lifetimes\cite{Pike2025, Quadir2024a, Hautzinger2025, Kassa2025}. Here, we report on the new photoabsorber CaCd$_2$P$_2$ after an early experimental screening that identified its particularly appealing balance of stability, relatively low bandgap, and co-catalyst compatibility.

\begin{figure}[h]
\centering
\includegraphics[width=\textwidth]{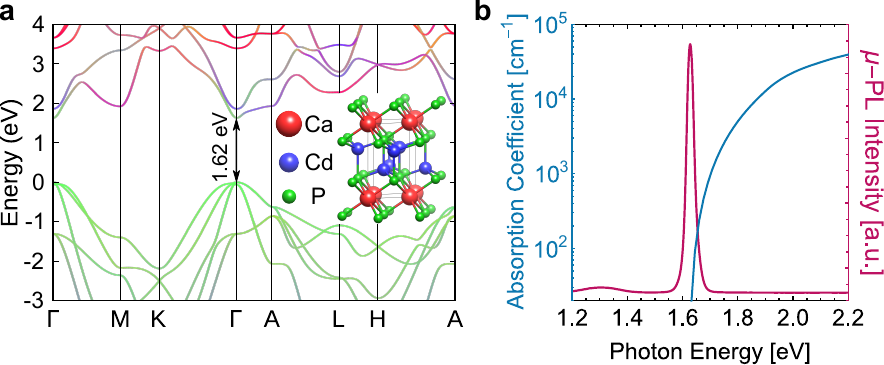}
\caption{CaCd$_2$P$_2$ Computation and Optoelectronic Characterization of CaCd$_2$P$_2$ Powder. a) Computed electronic band structure of CaCd$_2$P$_2$ with the crystal structure inset. The bands are color coded according to the elemental character of the electron states. b) The 0 K computed optical absorption coefficient of CaCd$_2$P$_2$ (log scale) alongside its 80 K $\mu$-PL emission spectrum (linear scale), showing strong emission at the predicted bandgap value. The band structure and absorption coefficient are calculated using the HSE06 hybrid functional\cite{Heyd2003, Heyd2006}. 
}
\label{basics}
\end{figure}

Fig. \ref{basics}a shows the electronic band structure of CaCd$_2$P$_2$,
revealing a direct bandgap of 1.62 eV and dispersive band edges. The color coding in Fig. \ref{basics}a indicates the elemental character of the electron states.
P is the dominant contributor to the top of the valence band, while hybridization of P orbitals with Ca and Cd orbitals plays a more significant role throughout the conduction band.  
Of particular note, 
the delocalized states at the band edges (i.e. those most responsible for the transport of electrons and holes) arise mainly from the atomic orbitals of P, and consequently are spatially concentrated around the P sublattice. The density of states with elemental decomposition, as well as real-space charge-density isosurfaces for the band extrema, can be found in Supplementary Fig. 1.  
However, many other conduction-band valleys have a mixed character, so photoexcited electrons will have the chance of concentrating in the other sublattices as they relax to the conduction band edge. The computed optical absorption coefficient indicates strong visible-light absorption (Fig. \ref{basics}b, Supplementary Fig. 2).

CaCd$_2$P$_2$ was synthesized by solid-state reactions, producing a powder with a wide range of particle sizes (from approximately 100 $\mu$m to sub-micron, by optical microscopy) with no secondary phases detected by X-ray diffraction (XRD) (Supplementary Fig. 3). This powder was used for investigation of the optoelectronic and photoelectrochemical properties.

Micro-photoluminescence ($\mu$-PL) measurements of the CaCd$_2$P$_2$ powder display a strong and narrow peak at approximately 1.6 eV (Fig. \ref{basics}b), which can be attributed to the band-to-band recombination. Additional examination of the temperature and power dependence of the emission corroborate the identity of this peak as arising from band-to-band recombination (Supplementary Fig. 4).

\section{Stability Under Oxygen-Evolution Conditions}

To assess the performance of CaCd$_2$P$_2$ as an OER photoelectrode material, we performed electrochemical measurements in 0.1 M KOH under 1-sun Air Mass 1.5 (AM1.5) illumination and in the dark. Working electrodes were prepared on glassy carbon (GC) by dropcasting ink comprised of the CaCd$_2$P$_2$ powder, acetylene black, and Nafion. Critically, no particular measures were taken to protect the semiconductor from the electrochemical environment.

\begin{figure}[h]
\centering
\includegraphics[width=\textwidth]{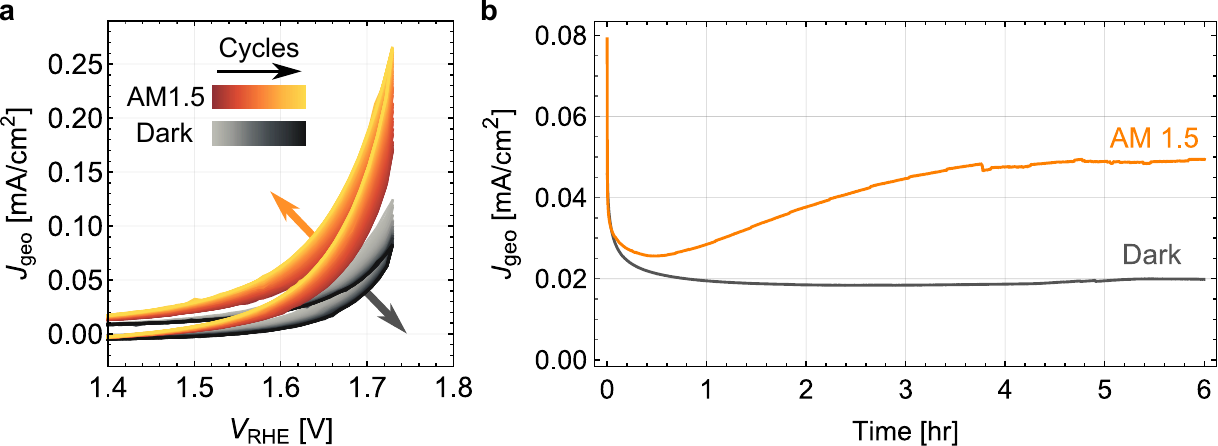}
\caption{Photoelectrochemistry of Bare CaCd$_2$P$_2$. a) CV of CaCd$_2$P$_2$ in OER showing the evolution of the sample under AM1.5 illumination and in the dark. Both sets of CV cycles were carried out until the change between subsequent cycles was negligible, 99 and 70 cycles for AM1.5 and dark conditions respectively. b) CA of CaCd$_2$P$_2$ at 1.7 V$_\text{RHE}$ under AM1.5 and dark illumination conditions, showing the activation period under illumination.
}
\label{PEC-neat}
\end{figure}

Cyclic voltammetry (CV) and chronoamperometry (CA) at 1.7 V$_\text{RHE}$ display a modest increase in current under illumination over the course of cycling or time, characteristic of an activation process (Fig. \ref{PEC-neat}a,b). In contrast, when cycled or held potentiostatically in the dark, no activation is seen. When removed from the light, and subjected to the same electrochemical conditions in the dark, the material was observed to undergo a deactivation process, as shown in the CV. The activation or deactivation periods were typically a few hours or less before stabilizing, depending on the experimental conditions, and if a deactivated sample was re-exposed to AM1.5 illumination, the prior activation could be recovered (Supplementary Fig. 5). To isolate the role of illumination on activation, CA measurements where the light was turned off (after illuminated activation) and the applied voltage was simultaneously increased by a quantity corresponding to the removed photovoltage still showed a decay in current (Supplementary Fig. 6).
Taken together, these results suggested that the light itself is playing a role in forming a meta-stable phase or phases at the semiconductor-electrolyte interface. 

Morphologically, the stabilizing effect of light on CaCd$_2$P$_2$ is clearly supported by scanning electron microscopy (SEM). Samples subjected to 140 hrs of CA at 1.6 V$_\text{RHE}$ under AM1.5 illumination or in the dark were compared against a freshly drop-cast control. Significant pitting and rounding of edges were consistently observed for the powder particles electrolyzed in the dark, while those subjected to AM1.5 illumination bore few to no obvious differences from those of the control, as shown in Fig. \ref{stability}a. More sample SEMs, including wider area, can be found in Supplementary Fig. 7-9.

We next sought to understand the nature of the light-stabilized interface. The short-lived nature of the AM1.5 activated surface 
unfortunately precluded X-ray photoelectron spectroscopy (XPS) measurement. Instead, we turned to wavelength-dependent Raman spectroscopy to examine the local bonding environments, using both a bulk-sensitive 514 nm laser and a more surface-sensitive 488 nm laser (which have predicted attenuation depths of 159 nm and 130 nm, respectively). By relying on the shallower absorption depth of higher energy photons, we sought to differentiate between contributions from the surface and the bulk.

\begin{figure}[h!]
\centering
\includegraphics[width=\textwidth]{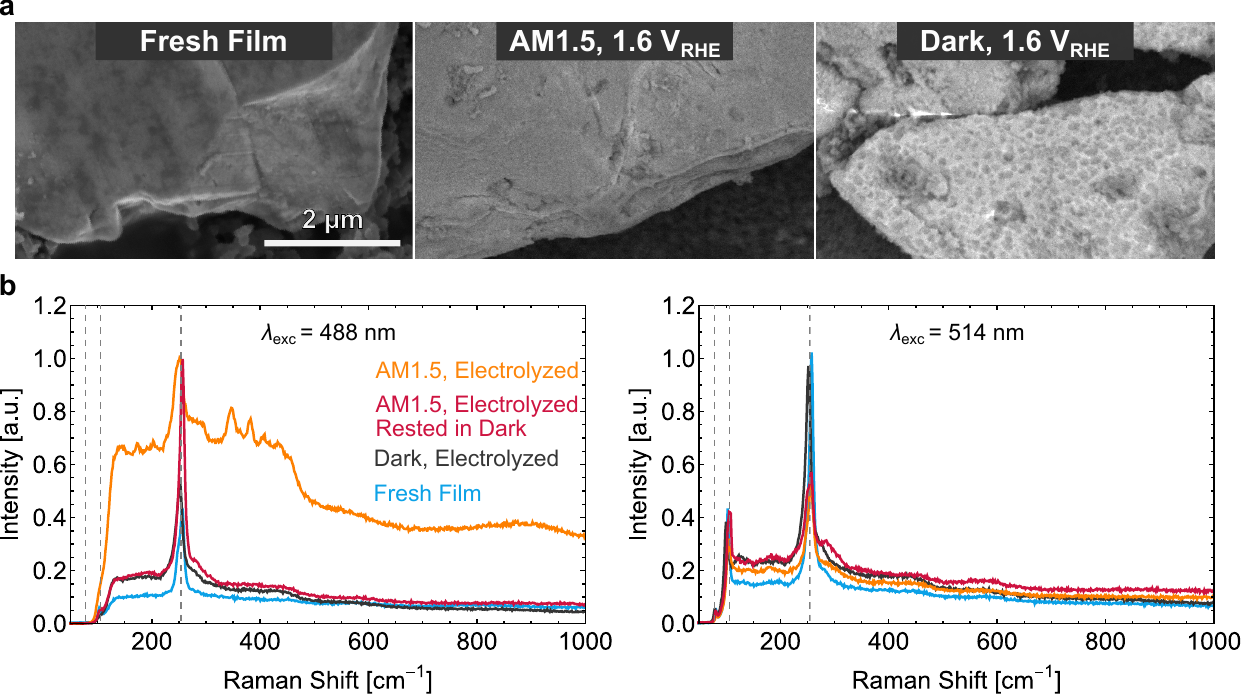}
\caption{Light-induced Stabilization of CaCd$_2$P$_2$. a) SEMs showing the surface morphology of CaCd$_2$P$_2$ particles in a freshly drop-cast electrode (left), and electrodes that have undergone electrolysis for 140 hr under AM1.5 (middle) and in the dark (right). Scale bar is shared. 
b) Wavelength dependent Raman spectroscopy of CaCd$_2$P$_2$ (electrolyzed for 48 hr) using 488 nm (left) and 514 nm (right) laser excitation. Vertical dashed lines are computed modes for CaCd$_2$P$_2$. The 488 nm spectra reveals the emergence of various new modes and a bright disordered background when CaCd$_2$P$_2$ undergoes electrolysis under AM1.5 vs in the dark. When the AM1.5 aged sample is allowed to rest in the dark, these features disappear. The general absence of such modes in the 514 nm spectra suggests that these transient phases are localized at the surface. All electrolysis for these panels was performed at 1.6 V$_\text{RHE}$ in 0.1 M KOH, and resting was done for 12 hrs at open-circuit conditions in 0.1 M KOH.}
\label{stability}
\end{figure}

Raman spectra were collected promptly following light-soaking and dark electrolysis (48 hr at 1.6 V$_\text{RHE}$) after removal from the electrolyte and power. Sample spectra for all conditions and excitation wavelengths are presented in Fig. \ref{stability}b. The spectra of powder particles on a fresh electrode (blue) cleanly matched that of computed modes (vertical dashed lines), especially under 514 nm light (right panel).  
When an electrode was subjected to electrolysis in the dark (dark gray), little difference was observed compared to the as-synthesized powder, save for a slightly brighter background. However, when an electrode was subjected to the same electrochemical conditions under AM1.5 illumination (orange), we observed the emergence of many additional Raman modes as well as a much brighter background, which were almost exclusively observed only when using the shorter wavelength light. It is worth noting that the rise in background is not attributable to leaking fluorescence, given the identical spectra collection across conditions. These collective features are suggestive of a complex surface, where non-CaCd$_2$P$_2$ phases are present alongside partial amorphization. When this same electrode was then allowed to rest in 0.1 M KOH in the dark, under open-circuit conditions, the spectra effectively reverted to the `dark' state. The full dataset can be found in Supplementary Fig. 10-11.

\section{CaCd$_2$P$_2$ Synergy with Co-based Co-catalyst}

\begin{figure}[h!]
\centering
\includegraphics[width=\textwidth]{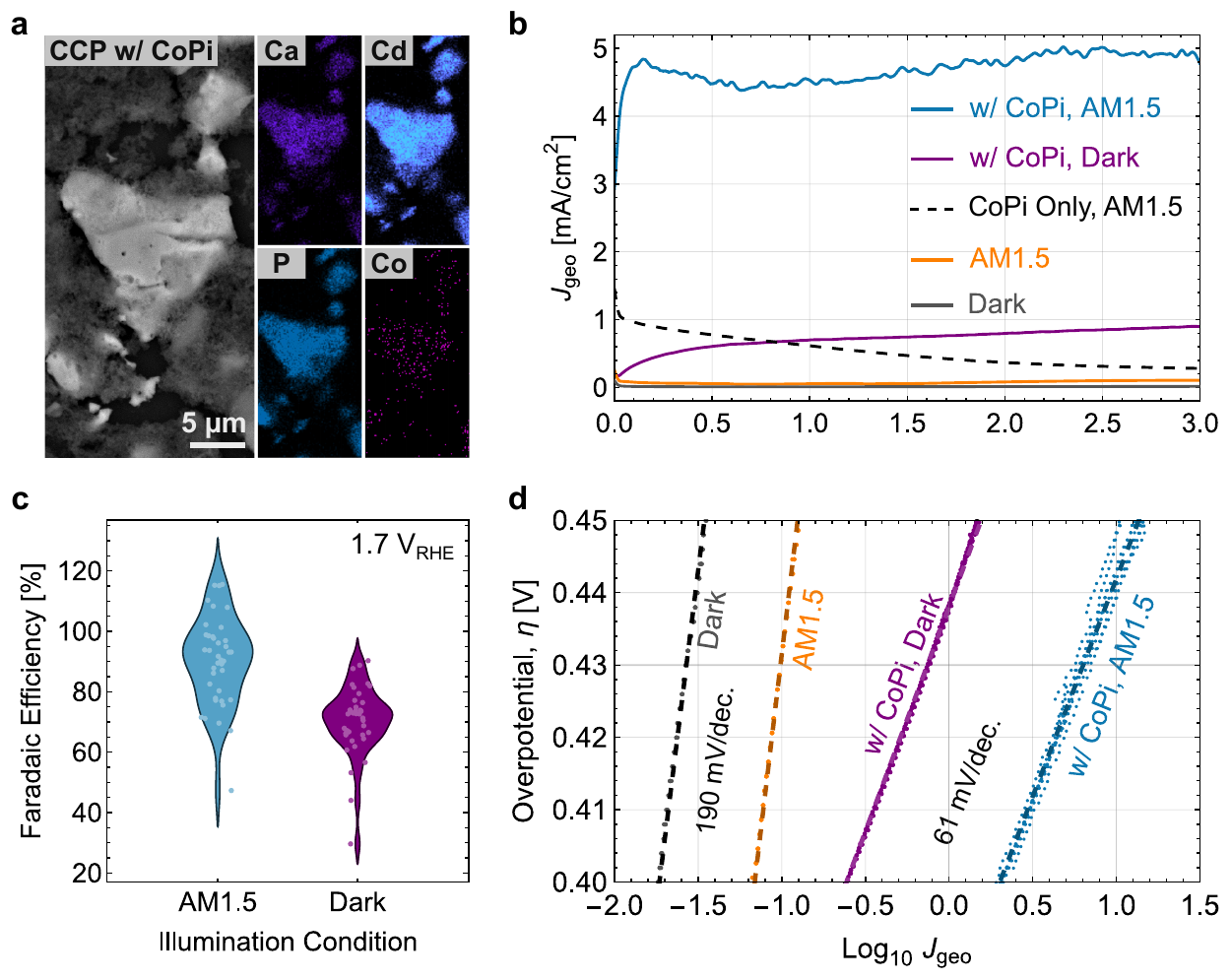}
\caption{CaCd$_2$P$_2$ Synergy with CoPi co-catalyst. a) SEM and EDS maps of a CCP electrode after CoPi deposition. Together, they show that the Co is concentrating on the CaCd$_2$P$_2$ particles. The brightness of the Co map was enhanced for clarity. b) CA of a single CaCd$_2$P$_2$ electrode with and without CoPi at 1.7 V$_\text{RHE}$ under AM1.5 and dark conditions, showing an increase in performance of two orders of magnitude, and generally stable current over the course of each measurement. When only CoPi is deposited (no CaCd$_2$P$_2$), we see clear degradation in performance, as expected in pure KOH electrolyte. c) Faradaic efficiency of CaCd$_2$P$_2$ with CoPi at 1.7 V$_\text{RHE}$ for 12 hrs under AM1.5 and dark conditions, showing enhanced efficiency under illumination. Sample collection was performed on a 20 minute cycle. d) Tafel analysis from CV comparing CaCd$_2$P$_2$ with and without CoPi, revealing significant improvements in onset voltage and Tafel slope. The electrolyte used for frames b) - d) was 0.1 M KOH. }
\label{copi}
\end{figure}

Despite the unusual stability displayed by CaCd$_2$P$_2$ under OER conditions, the photovoltages and currents that we observed are modest. We therefore sought to explore its compatibility with a co-catalyst and settled on Co-based CoPi. ``CoPi" collectively refers to a family of typically amorphous phosphate Co-based catalysts and related compounds\cite{Kanan2008, Kanan2009, McCrory2013, Vensaus2024}. CoPi is often formed \textit{in situ} and is well known for its self-healing abilities when utilized in a phosphate buffer, allowing it to replenish lost phosphate and cobalt during operation\cite{Thorarinsdottir2022}. In our studies we electrodeposited a small amount of CoPi (2.55 mC/cm$^2_\text{geo}$) on our electrodes before subjecting them to photoelectrolysis. We found significant synergy between CaCd$_2$P$_2$ and the deposited co-catalyst in both stability and overall performance. Energy dispersive X-ray spectroscopy (EDS) maps showing the concentration of Co at the semiconductor particles (versus other components of the drop cast film) are shown in Fig. \ref{copi}a. 

\FloatBarrier

A 1-2 order-of-magnitude improvement in current during OER (Fig. \ref{copi}b) and significantly lower overpotential was typical of the co-catalyst system as compared to neat CaCd$_2$P$_2$, as evidenced by Tafel analysis of CaCd$_2$P$_2$ with and without CoPi  (Fig. \ref{copi}d, Supplementary Table 1). Interestingly, the enhanced stability we observe for CaCd$_2$P$_2$ under photoelectrochemical OER conditions extends to the co-catalyst system as well. Faradaic efficiency over the 12 hours after activation under illumination is $97.4 \pm 2.8\% $), indicating a quantitative yield within the error of the measurement. In the dark, the Faradaic efficiency is only $67.2 \pm 2.3\%$.  When including the activation period, the FE under illumination is $90.2 \pm 1.6\%$ and in the dark only $70.4 \pm 1.5\%$. 
Additionally, chronoamperometry of the co-catalyst system has displayed stable and enhanced currents over 20+ hr measurements (Supplementary Fig. 13), whereas deposition of CoPi directly on glassy carbon shows a steady decline in performance (Fig. \ref{copi}b), as would be expected in a pure KOH electrolyte, though a high pH is thought to promote self-healing when phosphate is present\cite{Thorarinsdottir2022}. \section{Possible Mechanisms of Stability}

The origin of the light-induced stabilization of CaCd$_2$P$_2$ under OER conditions is of primary interest. Revisiting the electronic structure of Fig. \ref{basics}a, we note that both band extrema have strong P character, which implies that, at least for the bare semiconductor, reactivity and catalysis may be concentrated at surface P sites. In particular, the overwhelmingly P elemental character of the valence bands implies that as holes move to the surface to perform OER, they do so concentrated in the P sublattice, and thus may transit at P surface sites. Critically, this concentration might shield the metal cations from oxidation by these same holes, instead promoting the formation of oxidized phosphorus species such as PO$_x$ at the surface. XPS of the native surface of CaCd$_2$P$_2$ displays a rich chemistry (Supplementary Fig. 14-15). While phosphates and carbonates are abundant, the presence of oxides is relatively muted, which could be explained in part by the enhancement in reactivity of the P sites evidenced by the electronic band structure of CaCd$_2$P$_2$. 

Pseudo-external quantum efficiency measurements (unknown total absorption) of a fresh electrode in electrolyte show a distinct onset of current flow without external bias (Supplementary Fig. 16). This onset occurs close to the semiconductor bandgap and increases with increasing photon energy, likely due to increased carrier generation near the surface, supporting the hypothesis of an activation process mediated by band-edge absorption.

These data indicate a light-driven surface reaction. Presently, we can tentatively infer from the surface and Raman spectroscopy (Supplementary Table 2) that Ca- and Cd-(oxy)hydroxides appear to be some of the most likely candidates for the light stabilized phases. \textit{Operando} spectroscopy of the surface during activation and bulk electrolysis may be a future path to determine the surface chemistry more conclusively.

The bonding character of the band edge states is also of importance. In many conventional low-bandgap semiconductors, such as InP (E$_g = 1.34$ eV), it is believed that the valence and conduction bands are comprised of bonding and antibonding orbitals, respectively. Consequently, when an electron absorbs a photon and transitions between these bands, the cohesive energy of the lattice decreases. In these materials, the very act of absorbing a photon weakens the bonds. In addition to self-oxidation by holes, this weakening of the bond strength is thought to be one of the key mechanisms by which these semiconductor photoelectrodes degrade under OER conditions\cite{Su2017}.  

This mechanism of light-induced oxidative corrosion seen in traditional semiconductors is in direct contrast to the case of CaCd$_2$P$_2$. By inspection of the charge density isosurfaces (Supplementary Fig. 1b), it can be seen that neither of the edge states have the character of a bonding orbital, as this would concentrate charge density in the interatomic space rather than only around specific atoms. Furthermore, energy-resolved analysis of the bonding interaction seems to confirm that both of these states have an antibonding character (Supplementary Fig. 17). The antibonding character of both bands may be an additional mechanism by which CaCd$_2$P$_2$ may be protected from photocorrosion.

\section{Conclusions}

Starting from first-principles computation, we have identified the ternary Zintl phosphide, CaCd$_2$P$_2$, as a 1.6 eV bandgap semiconductor of interest as a photoabsorber for water oxidation. The neat semiconductor powder displays exceptionally stable performance in OER conditions for a semiconductor of its bandgap. Its electrochemical stability can be attributed in part to surface transformation under AM1.5 illumination. This transformation imparts improved chemical and morphological stability and is shown to be comprised of a highly complex surface, including a mixture of amorphous and ordered components, which dissipate when the semiconductor is returned to the dark. Furthermore, the active CaCd$_2$P$_2$  surface readily synergizes with electrodeposited CoPi, and the stable co-catalyst system shows both significant improvements in Tafel slope and reduced overpotential.

Taken together, the computational and experimental data on CaCd$_2$P$_2$ highlight several characteristics that may be extendable to other materials for the design of stable photoelectrochemical systems. The P-dominated valence band and antibonding character at the band edges of CaCd$_2$P$_2$ may possibly reduce susceptibility to direct photocorrosion. In addition, CaCd$_2$P$_2$ develops a semiconductor-electrolyte interphase layer, in a limited analogy to the solid-electrolyte interphase that develops in Li ion batteries that provides graphite electrode passivation\cite{Peled1979, Aurbach1987,Fong1990, Luo2021, Gu2023,SutingWeng2023, Wang2018a,Aktekin2023}. This interphase layer endows CaCd$_2$P$_2$ with unique stability among semiconductors of similar bandgap in alkaline OER conditions. 

Encouragingly, CaCd$_2$P$_2$ is only one case study among the AM$_2$P$_2$ family, many of which are promising photoabsorbers\cite{Yuan2024,Quadir2024a,Hautzinger2025,Pike2025, Kassa2025}. Leveraging more broadly the active formation of a stable, passivating interphase layer holds promise to both grow the material set available for water oxidation and to reduce the complexity required to realize stable solar fuel production.

\section{Methods}

\subsection{Computational Details}
The detailed procedure for the high-throughput computational materials screening can be found in the work of Yuan et al\cite{Yuan2024}. That work marked the discovery the of the AM$_2$P$_2$ family as promising light-absorbing semiconductors. Many of these Zintl phases (A = Ca, Sr, Ba, M = Zn, Cd, Mg)  have bandgaps in the range of $\sim$1.5–2.0 eV \cite{Pike2025}.

In the present work on CaCd$_2$P$_2$, first-principles density-functional theory (DFT) calculations were performed using the VASP code\cite{Joubert1999, Vargas-Hernandez2020}. The standard
VASP PAW pseudopotentials (PBE version 54) were used: Ca$\_$sv, Cd, and P. A
plane-wave basis set with a cutoff energy of 400 eV was used
to expand electron wave functions. The HSE06 hybrid functional\cite{Heyd2003, Heyd2006} was used to calculate the equilibrium crystal structure, electronic band structure, and optical absorption coefficient for CaCd$_2$P$_2$. Using a $8\times8\times5$ $\mathbf{k}$-point mesh for Brillouin-zone integration, the lattice constants of CaCd$_2$P$_2$ were found to be: $a=b=4.289$ \AA and $c=7.036$ \AA . To determine the optical absorption coefficient, the frequency-dependent dielectric matrix was computed within the independent-particle picture\cite{Gajdos2006} and with a $12\times12\times6$ $\mathbf{k}$-point mesh; for determination of the real part of the dielectric tensor using the Kramers-Kronig transformation, a small complex shift of $10^{-5}$ was used. The Python toolkit Sumo\cite{MGanose2018} was used to plot and analyze the electronic band structure and optical absorption spectra. Crystal orbital Hamilton populations (COHP) analysis\cite{Dronskowski1993} was performed using the Lobster\cite{Maintz2016, Nelson2020}, and Lobsterpy\cite{Naik2024} packages.

For the calculation of the Raman modes, the crystal structure was relaxed with the PBEsol functional, resulting in the lattice constants of $a=b=4.249$ \AA  and $c=6.921$ \AA. The irreducible representation of the phonon modes at Brillouin-zone center in the trigonal \textit{P}$\bar{3}$\textit{m}1 structure is $\Gamma=2\mathrm{A}_\mathrm{1g}+2\mathrm{E}_\mathrm{g}+2\mathrm{A}_\mathrm{2u}+2\mathrm{E}_\mathrm{u}$, where the $\mathrm{A}_\mathrm{1g}$ and $\mathrm{E}_\mathrm{g}$ modes are Raman active\cite{Homes2023}. The Python software toolkits Phonopy\cite{Togo2023} and Phonopy-Spectroscopy\cite{Skelton2017} were used to postprocess phonon calculations and extract the frequencies of the Raman modes.

\subsection{Materials}

CaCd$_2$P$_2$ was synthesized from elemental Ca (99.98\%, Alfa Aesar), Cd (99.95\%, Alfa
Aesar), and red phosphorus (98.90\%, Alfa Aesar). Anhydrous CoCl$_2$ was synthesized from elemental Co (99.8\%, MilliporeSigma) and HCl (ACS Plus Certified, Fisher Chemical), with isopropyl alcohol (HPLC Grade, Fisher Chemical) used for solvent exchange. Electrolyte preparation was performed with Milli-Q ultrapure water, KOH (ACS Certified, Fisher Chemical), KH$_2$PO$_4$ (ACS Certified, Fisher Chemical), and K$_2$HPO$_4$ (ACS Reagent, $\geq$ 98\%, MilliporeSigma). Electrode preparation was done with Nafion 5\% wt. solution in aliphatic alcohols and water (MilliporeSigma) and acetylene black suspended in Milli-Q water and ethanol (Anhydrous, USP Specs, KOPTEC).

\subsection{CaCd$_2$P$_2$ Synthesis}

The elements in the stoichiometric
1:2:2 ratio (Ca:Cd:P) were placed into a carbonized 9/11 mm inner/outer diameter silica ampoule in Ar atmosphere of the glovebox. The ampoule was evacuated and then sealed
using hydrogen-oxygen torch. The sealed ampoule was heated in a muffle furnace to a
temperature of 1000 $^\circ$C in 8 hours and annealed at that temperature for 48 hours. After
cooling in the turned-off furnace, the ampoule was then opened in ambient conditions.
The resulting sample was ground in an agate mortar and reannealed at 800 $^\circ$C for 48
hours in an evacuated carbonized silica ampoule. The product obtained afterwards was determined to be of single-phase according to powder X-ray diffraction, as measured using a Rigaku MiniFlex 600 benchtop diffractometer with Cu-\textit{K}$\alpha$ radiation ($\lambda$ = 1.5406 ) and Ni-\textit{K}$_\beta$ filter in 3$^\circ \leq 2\theta \leq 90^\circ$ range with 0.02$^\circ$ steps
at 10$^\circ$/min. The experimental pattern is compared to the calculated one obtained from the published crystal structure.

\subsection{Photoelectrochemical Measurements}

All electrochemical measurements were performed with a VSP-300 Potentiostat (BioLogic). Unless otherwise noted, photoelectrochemical measurements were performed at room temperature in a borosilicate glass cell with a JGS2 optical quartz window (Dek Research), using a custom machined PTFE lid sealed with Viton o-rings. All measurements were performed in 0.1 M KOH solution. All measurements, except Faradaic efficiency, were performed with continuous N$_2$ purging with rotameter controlled flow of approximately 30 sccm. A submerged glass frit was used for all gases with stirring on. Working electrodes were prepared by dropcasting inks on glassy carbon. Measurements were performed with a graphite rod counter electrode (Pine Research), and Ag/AgCl single junction reference electrodes (Pine Research). Reference electrodes were regularly calibrated against a prepared standard hydrogen electrode. Simulated AM1.5 illumination was supplied by a Class AAA Pico Solar Simulator with a 350-1500 nm spectral range (G2V Optics Inc). To account for heating effects, the cell was submerged in water in a re-crystallization dish (up to just below the cell lid surface) to serve as a heat sink, which was topped off on a regular basis for the longest running experiments.

Faradaic efficiency measurements were performed with a MG\#5 Gas Chromatograph (SRI Instruments) with a MolSieve 5A and Mayssep D column, set with an oven temperature of 70 $^\circ$C and He pressure of 10 psig. Through the cell He was flowed as the carrier gas at a 5.000 sccm setpoint using a SmartTrak 100 mass flow controller (Sierra Instruments), and the head space (20m mL) was purged for at least 1 hr, during open circuit voltage (OCV) stabilization. Air was used as the calibration gas for oxygen peak area. 
 Gas sampling was done online. Oxygen was generated over the course of 12 hr CA, with gas sampling being performed every 20 minutes. To account for any background signal, background measurements were performed before and after OER when no oxygen generation was occurring. The ratio of O$_2$ and N$_2$ peak areas in the background measurement was calculated, and a time-dependent linear calibration fit was made to account for instrument drift. For the samples taken during OER, the area of the $N_2$ peak was used in conjunction with the calibration fit to subtract a proportional amount of any background O$_2$, leaving the area corresponding to the generated oxygen. For the calculation of Faradaic efficiency, the uncertainty was calculated by the standard error of the calibration fit and an assumed flow uncertainty of $\pm 0.02$ sccm. The Faradaic efficiency values reported after activation averaged the final 6 hours of measurement.

Before and after each use, all glassware and electrodes were triple rinsed with distilled water, followed by an additional rinse in Milli-Q water. All glassware was additionally regularly cleaned with 1 M HCl, followed by the water rinse above, and this was also performed after every set of CoPi based measurements.

A real-time iR compensation of 85\% was used for used for all measurements, with the uncompensated resistance being determined by PEIS, and no additional iR correction was applied in post processing.  
All reported CV was collected at a scan rate of 20 mV/s. Stabilized background CV scans were collected prior to film deposition and used for background correction of presented stabilized data. These same corrected and stabilized scans are what was used to determine the reported photovoltage and photocurrent of the neat CaCd$_2$P$_2$.

\subsection{CaCd$_2$P$_2$ Working Electrode Preparation}

All working electrodes used in this study were prepared on glassy carbon disks (5 mm diameter) held in an L-shaped PTFE housing (Dek Research). Glassy carbon preparation involved polishing the disks to a mirror finish using alumina polishing compound (5, 0.3, 0.05 $\mu$m) on microfiber pads with alternating circular motions. Alumina residue was removed by bath sonicating in Milli-Q water for 15 minutes, followed by overnight soaking in 1M HCl, and finished with another rinse with Milli-Q water and blown dry with filtered air from the house compressor.

Inks were prepared using a loading of 5 mg/mL of CaCd$_2$P$_2$ in a 2:3 EtOH:H$_2$O solvent mixture. Acetylene black was used as a conductive filler with a loading of 1 mg/mL. This mixture was probe sonicated for 30 minutes (20 s cycles with 50\% duty) in order to promote the initial dispersion. After probe sonication Nafion solution (5\% wt.) was added at 8 $\mu$L/mL for use as a binder. Once Nafion was added, the inks were bath sonicated for 15 min every time immediately before use.

Before drop casting the films, the glassy carbon electrodes were preheated on a hotplate with a setpoint of 100 $^\circ$C. Immediately after bath sonication 10 $\mu$L were drawn from the middle of the ink column and dropped on the preheated glassy carbon. Drying was allowed to proceed for 30 minutes, after which the hotplate was turned off and allowed to cool to room temperature before the electrode was removed.

Owing to challenges with the mechanical integrity of the films, some powder was ball-milled for 30 minutes using a dedicated zirconia container and media while suspended in ethanol. This milling served to reduce the size of the largest particles, but SEM confirmed that the apparent majority of the CaCd$_2$P$_2$ remained in the form of particles $>1\mu\text{m}$. We found this to significantly reduce delamination of the films but did not result in any notable enhancement in catalytic activity, as verified by ECSA corrected chronoamperometry. This milled CaCd$_2$P$_2$ was used in the Fig. \ref{PEC-neat}b and Fig. \ref{copi}c. All other data was collected with as-synthesized powder.

\subsection{Electrodeposition of CoPi}

The electrodeposition of CoPi was adapted from a procedure described elsewhere\cite{Vensaus2024}.

CoCl$_2$ was synthesized by reacting Co metal in excess HCl. After reaction and separation, the hydrated product was made anhydrous by solvent exchange with isopropyl alcohol, producing a blue powder that was stored in a sealed vial whenever not in use.

Electrodeposition of CoPi was done in a 5-neck electrochemical cell made of borosilicate glass (Pine Research) in N$_2$ purged 0.1 M K$_2$HPO$_4$/KH$_2$PO$_4$ buffer solution (pH 7). The electrodes used were a Pt wire counter electrode (Pine Research) and Ag/AgCl single junction reference electrode (Pine Research). 

To this electrolyte was added CoCl$_2$, forming a 0.4 mM solution. Due to the tendency to form visible precipitates within 30-60 minutes, the CoCl$_2$ was always added under vigorous stirring and at most 10 minutes prior to electrodeposition. This time window proved sufficient to allow for complete dissolution, based on visual inspection. Electrodeposition was done with chronoamperometry at 1.5 V$_\text{RHE}$, terminating when 0.5 mC of charge was passed. Additionally, the Faradaic efficiency measurements, and the data in Supplementary Fig. 17, were conducted with an additional etch of the drop cast film in 0.01 M HCl for 30 minutes immediately prior to CoPi deposition. This was done with the intent of creating a pristine surface for the CoPi to make intimate contact with, though the effect of this step, if any, is inconclusive.

\subsection{$\mu$-Photoluminescence Measurements}
Temperature and power dependent $\mu$-photoluminescence measurements were conducted on a Horiba Labram HR Evolution Raman spectrometer, using a 532 nm excitation laser while employing a 50x objective lens and 300 l/mm grating. The CaCd$_2$P$_2$ powder was pressed into a pellet for these measurements. The power dependent data were all collected from the same spot on one particle. The reported power densities are based on power measurements accounting for the optical efficiency of the system as well as neutral density filters employed in the measurements. An approximate spot size of 3 $\mu$m was determined by low intensity imaging on a polished surface.

\subsection{Raman Measurements}
Raman spectra were collected on a Renishaw inVia Raman spectrometer, with 488 and 514 nm excitation lasers with a 50x objective and a 2400 l/mm grating. The chronoamperometry measurements for the AM1.5 and dark conditions were carried out synchronously in separate cells. After the conclusion of chronoamperometry, the assembled cells were immediately transported to the Raman spectrometer. Once removed from the electrolyte, the working electrodes were placed directly on the microscope stage after a rinse with Milli-Q water and drying with compressed air. Spectra were collected as quickly as possible after removal from the electrolyte, with one spectrum per particle. In the case of the AM1.5 illuminated sample, a UV lamp was shone through the window on the cell during transit, with the intent of maintaining activation. This sample was placed back into the cell (in 0.1 M KOH) with no connected power source, and allowed to rest in the dark for 12 hours before collection of additional Raman spectra.

\subsection{Scanning Electron Microscopy and Energy Dispersive X-Ray Spectroscopy}
SEM and EDS micrographs were collected on a Zeiss Sigma 500 Scanning Electron Microscope. With the exception of Fig. \ref{copi}a, all SEM micrographs were collected using an inLens secondary electron detector, accelerating voltage of 3.00 kV, and an aperture of 30.00 $\mu$m. The SEM and EDS mapping in Fig. \ref{copi}a used an accelerating voltage of 15.00 kV, and an aperture of 30.00 $\mu$m. Mapping was concluded once 2M counts were reached.

\subsection{Optical Microscopy}
Optical microscopy images  
were taken on a VHX-X1 Series Digital Microscope (Keyence).

\subsection{Pseudo-External Quantum Efficiency}
The stabilized light from a Xenon discharge lamp (6258, Newport) was decomposed with the use of a monochromator (Cornerstone 260, Newport). Prior to entering the monochromator, the light is passed through one of several possible filters (335, 590, 1000 nm longpass, depending on the target wavelength) in order to minimize the signal from grating harmonics. The resulting monochromator output was chopped using a mechanical chopper and focused using a set of lenses. The power of this focused spot was measured using a photodiode/power meter (S120VC, Thorlabs) across a range of wavelengths with a 5nm step size. The power meter was then swapped out for an assembled electrochemical cell, with the focused light incident on the working electrode surface, with the spot being visually contained within the electrode area. The working and counter electrodes were connected to a current-to-voltage pre-amplifier (SR570, Stanford Research Systems), with no bias applied by the pre-amplifier. The output of this pre-amplifier was then routed to a lock-in-amplifier (SR830, Stanford Research Systems) using the optical chopper for reference, and the output voltage of the lock-in amplifier was recorded for the same wavelengths and step size. This output voltage was then scaled according to the number of photons computed for each step, according to the power meter measurement, and the resultant data is presented. The code used to automate this measurement can be found on the Fenning Research Group \textit{GitHub} repository. 

\subsection{X-Ray Photoelectron Spectroscopy}
X-ray photoelectron spectroscopy (XPS) was performed using an AXIS-Supra by Kratos Analytical with Au and Cu calibration, using a Al K-$\alpha$ photon source. Survey and high-resolution scans of Ca 2p, Cd 2p, P 2p, C 1s, and O 1s were performed on the as-synthesized powder mounted on Cu tape, with the electron flood gun in use to counteract charging. Correction for the effect of the flood gun was done with a C 1s adventitious carbon offset. XPS peak fitting was done in CasaXPS using Shirley backgrounds. Lorenztian asymmetric lineshapes were used to fit the peaks, with scans to optimize the lineshape parameters. Fitting constraints of area, full-width-half-max, and doublet separation\cite{Moulder1992} were used. Identification of Ca$_3$(PO$_4$)$_2$ was based on spectra with adventitious carbon calibration\cite{Barbaux1992}. Identification of Cd$_3$(PO$_4$)$_2$ was based on spectra with no stated calibration method\cite{Lee2018}. Identification for carbonates and oxides were based on peak positions listed in the NIST XPS database, and were exclusively limited to spectra collected with the use of Au instrument calibration\cite{Sosulnikov1992a, Stipp1991a, Landis2016, Hammond1975}.

\subsection{Experimental Data Analysis}
Other than XPS fitting, all experimental data analysis was performed on \textit{Wolfram Mathematica 12.3} using custom-built code. Various functions were improved or built with the use of \textit{ChatGPT 4o}.

\section{Contributions}

G.L.E. co-designed experiments, undertook data collection and analysis, and was the primary writer of the manuscript. Z.Y. co-designed the computational screening methodology, conducted first-principle calculations, and co-wrote the manuscript. M.R.H. co-designed and carried out the material synthesis. Y.C. conducted first principle calculations and co-wrote the manuscript. G.K. undertook $\mu$-PL data collection and provided manuscript feedback. V.S.D. assisted in early experimental screening and provided manuscript feedback. T.N. assisted in Faradaic efficiency measurements and provided manuscript feedback. J.R.P. assisted in pseudo-external quantum efficiency measurements and provided manuscript feedback. J.L. supervised the project and provided manuscript feedback. K.K. supervised the project, co-designed the material synthesis, and provided-manuscript feedback. G.H. supervised the project, co-designed the computational screening methodology, and provided manuscript feedback. D.P.F. supervised the project, co-designed experiments, and co-wrote the manuscript.

\section{Acknowledgements}
This work was supported by the U.S. Department of Energy, Office of Science, Basic Energy Sciences, Division of Materials Science and Engineering, Physical Behavior of Materials program under award number DE-SC0023509. This research used resources of the National Energy Research Scientific Computing Center (NERSC), a DOE Office of Science User Facility supported by the Office of Science of the U.S. Department of Energy under contract no. DE-AC02-05CH11231 using NERSC award BES-ERCAP0023830. The authors acknowledge the use of facilities and instrumentation at the UC Irvine Materials Research Institute (IMRI), which is supported in part by the National Science Foundation through the UC Irvine Materials Research Science and Engineering Center (DMR- 2011967). Thanks to Dr. Ich Tran for support with photoelectron spectroscopy data collection. This work was performed in part at the San Diego Nanotechnology Infrastructure (SDNI) of UC San Diego, a member of the National Nanotechnology Coordinated Infrastructure, which is supported by the National Science Foundation (grant ECCS-2025752).

\bibliography{export}

\end{document}


\newpage 

\tableofcontents

\newpage

\section{Supplementary Figures}

\begin{figure}[h!]
\centering
\includegraphics[width=0.95\textwidth]{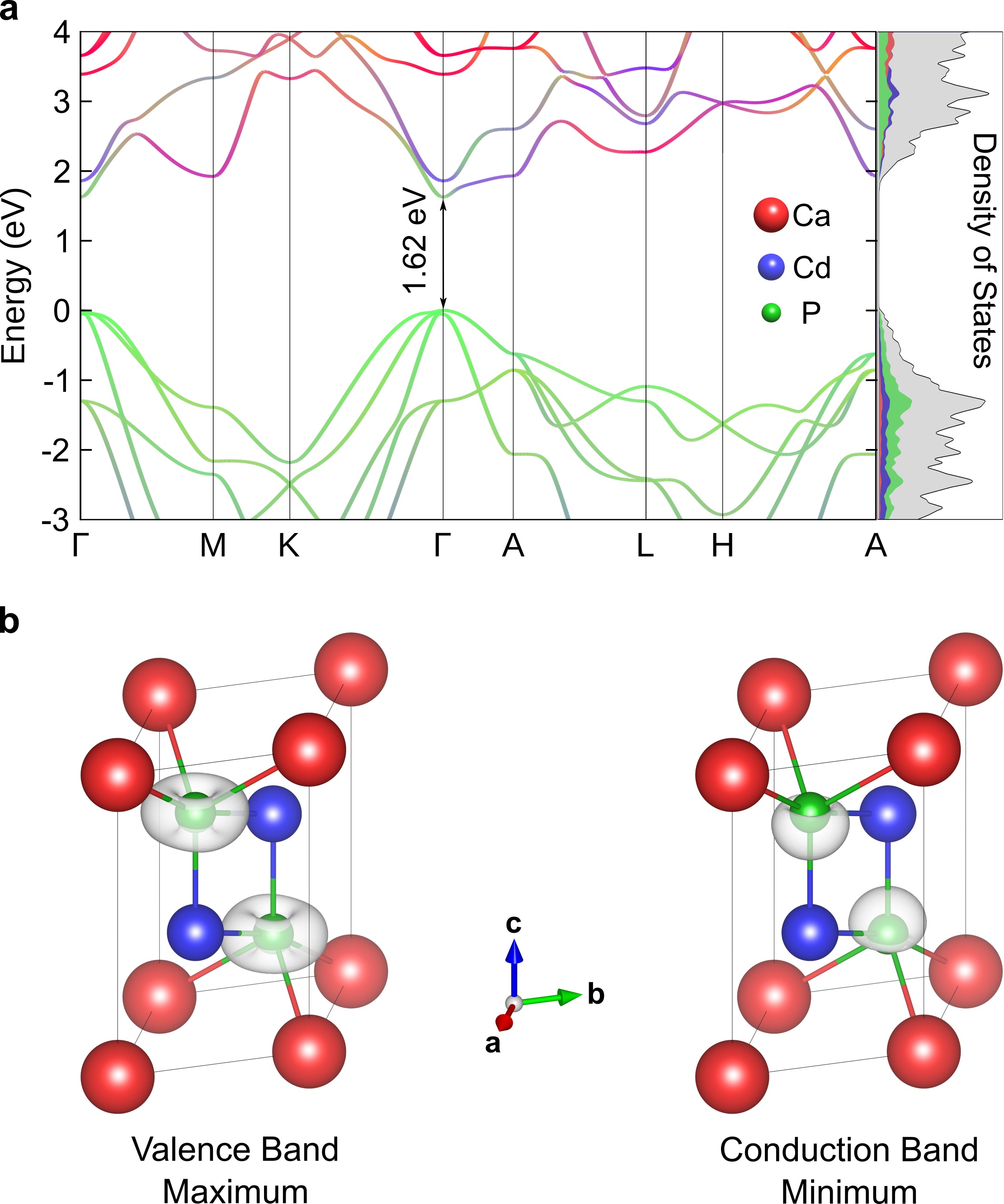}
\caption{
Electronic States of CaCd$_2$P$_2$. a) The electronic band structure of CaCd$_2$P$_2$ (CCP) color coded by elemental character (left) plotted alongside the density of states (right). b) Charge density isosurfaces (white volume) of the CCP band extrema, the valence band maximum (left) and the conduction band minimum (right). Both show a distinct concentration of charge density around the P sublattice. Computation was performed with the HSE06 hybrid functional.
}
\end{figure}

\newpage

\begin{figure}[h!]
\centering
\includegraphics[width=0.8\textwidth]{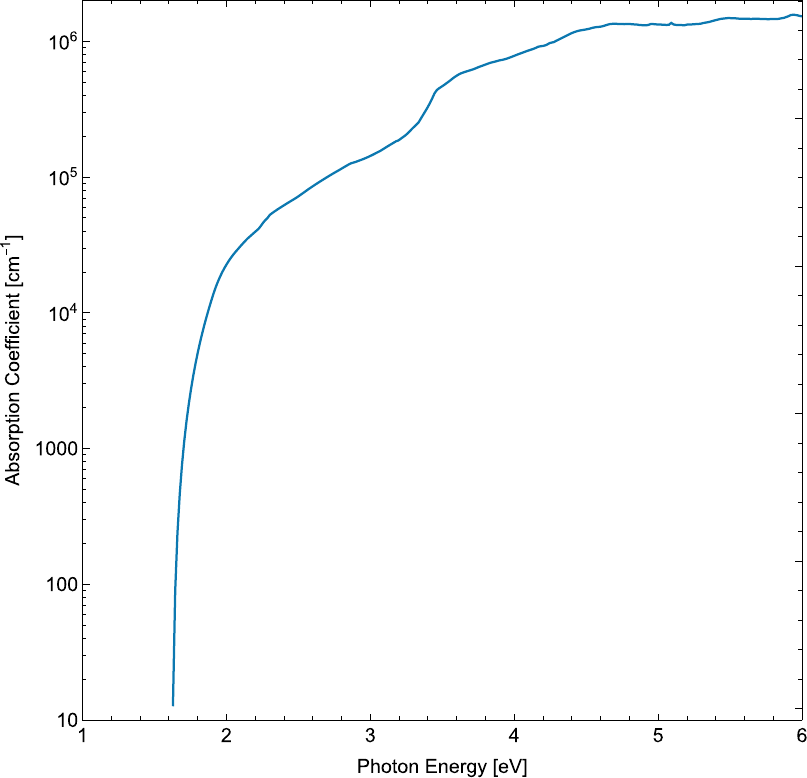}
\caption{
Computed Absorption Coefficients of CaCd$_2$P$_2$.
}
\end{figure}

\newpage

\begin{figure}[h!]
\centering
\includegraphics[width=0.85\textwidth]{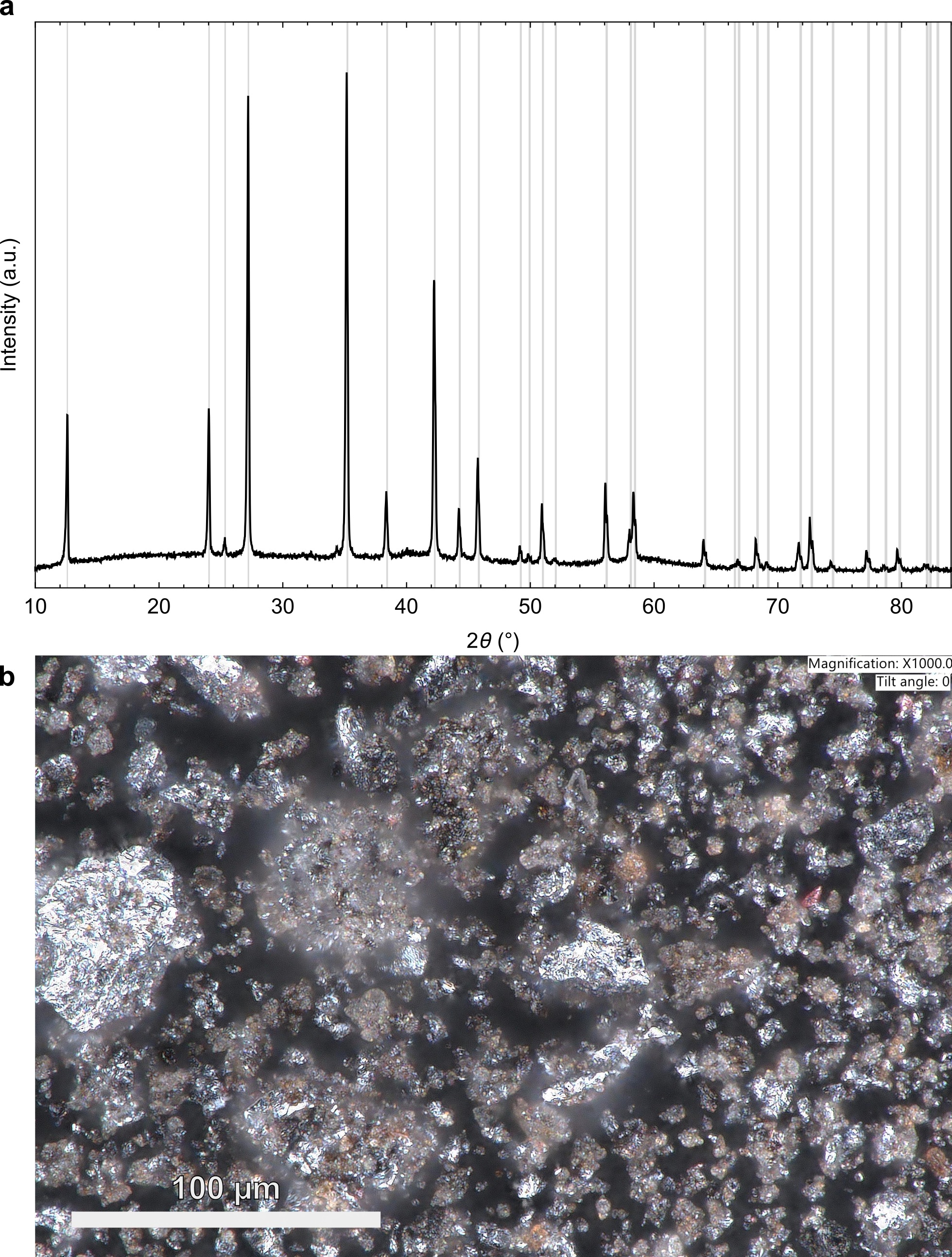}
\caption{As-synthesized Powder. a) X-ray diffraction pattern of the as-synthesized powder, compared against computed peaks (vertical lines), showing no secondary bulk phases. b) Optical microscopy image of the as-synthesized CaCd$_2$P$_2$ powder.
}
\end{figure}

\newpage

\begin{figure}[h!]
\centering
\includegraphics[width=\textwidth]{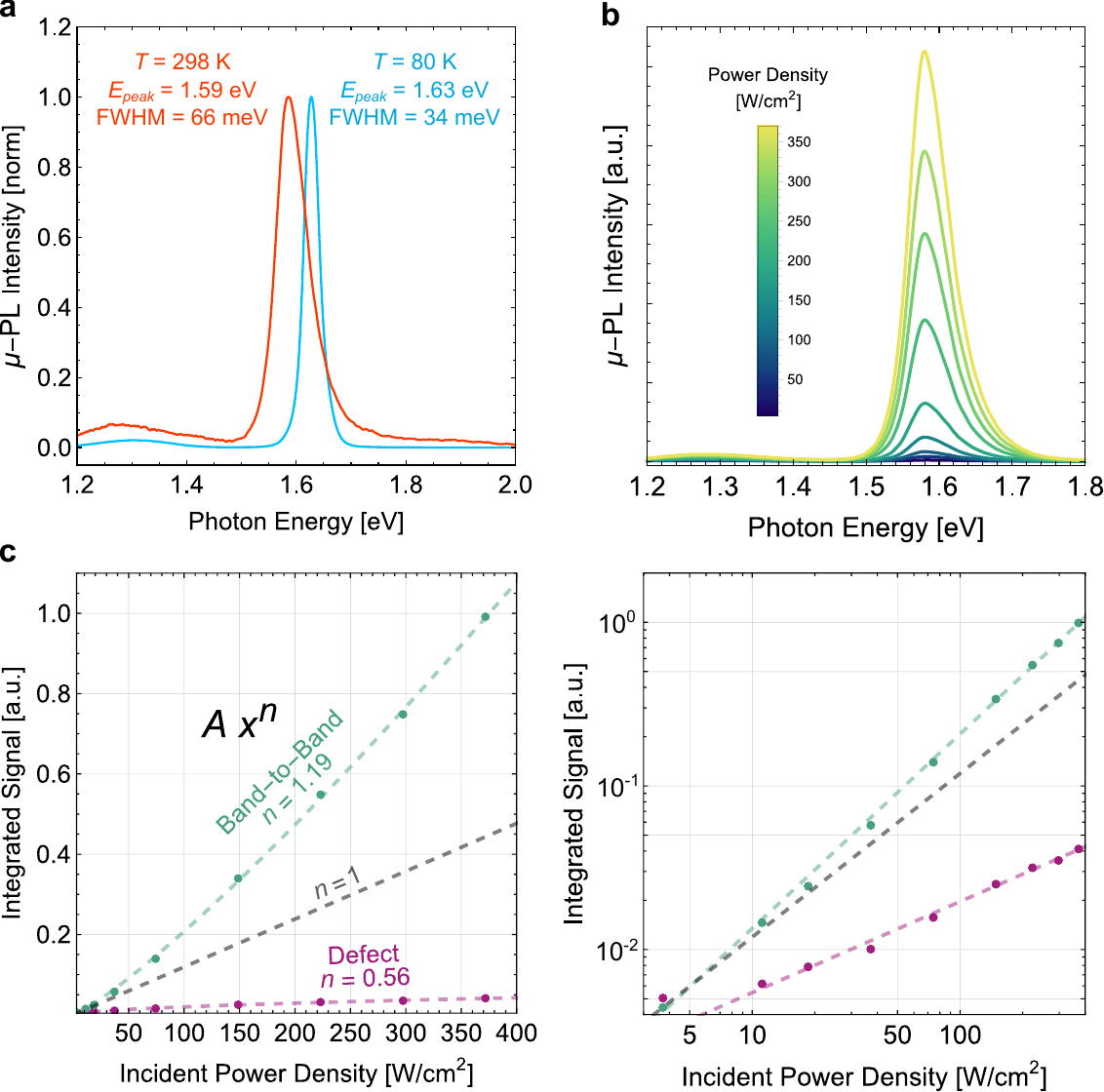}
\caption{Band-to-band Photoluminesce. a) $\mu$-Photoluminescence ($\mu$-PL) spectra of CCP at 80 K (blue) and 298 K (red). The primary peak undergoes the expected blue-shift as the temperature is lowered and the lattice contracts, as well as decrease in thermal broadening from the shortening of the Fermi tails. b) Power-dependent $\mu$-PL spectra taken at one particle location at 298 K. The emission of a luminescent defect is visible at lower energies. c) Integrated power measurements from the data in b) in linear scale (left) and loglog scale (right), fitted against $A x^n$. The band-to-band transition shows a subquadratic relationship ($n = 1.19$) and the defect shows a sublinear relationship ($n = 0.56$). 
}
\end{figure}

\newpage

\begin{figure}[h!]
\centering
\includegraphics[width=\textwidth]{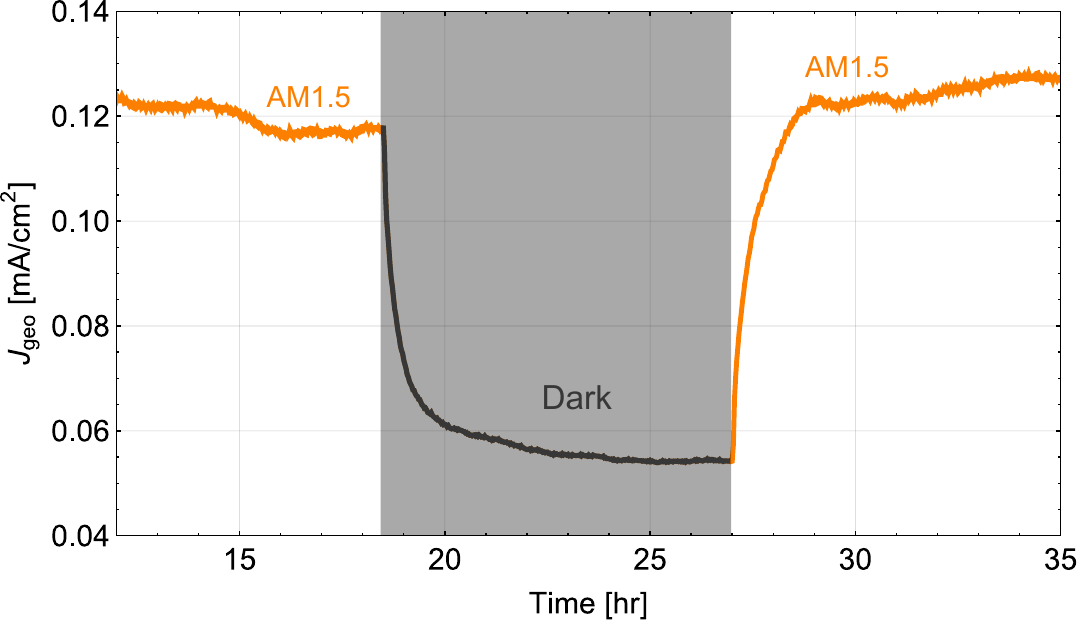}
\caption{Chronoamperometry (CA) of a CCP electrode at 1.7 V$_\text{RHE}$, when AM1.5 illumination is removed, resulting in deactivation, and reactivation once illumination is returned. Temperature control for this measurement
was done with a hotplate with temperature probe (T = 35 $^\circ$C).
}
\end{figure}

\newpage

\begin{figure}[h!]
\centering
\includegraphics[width=\textwidth]{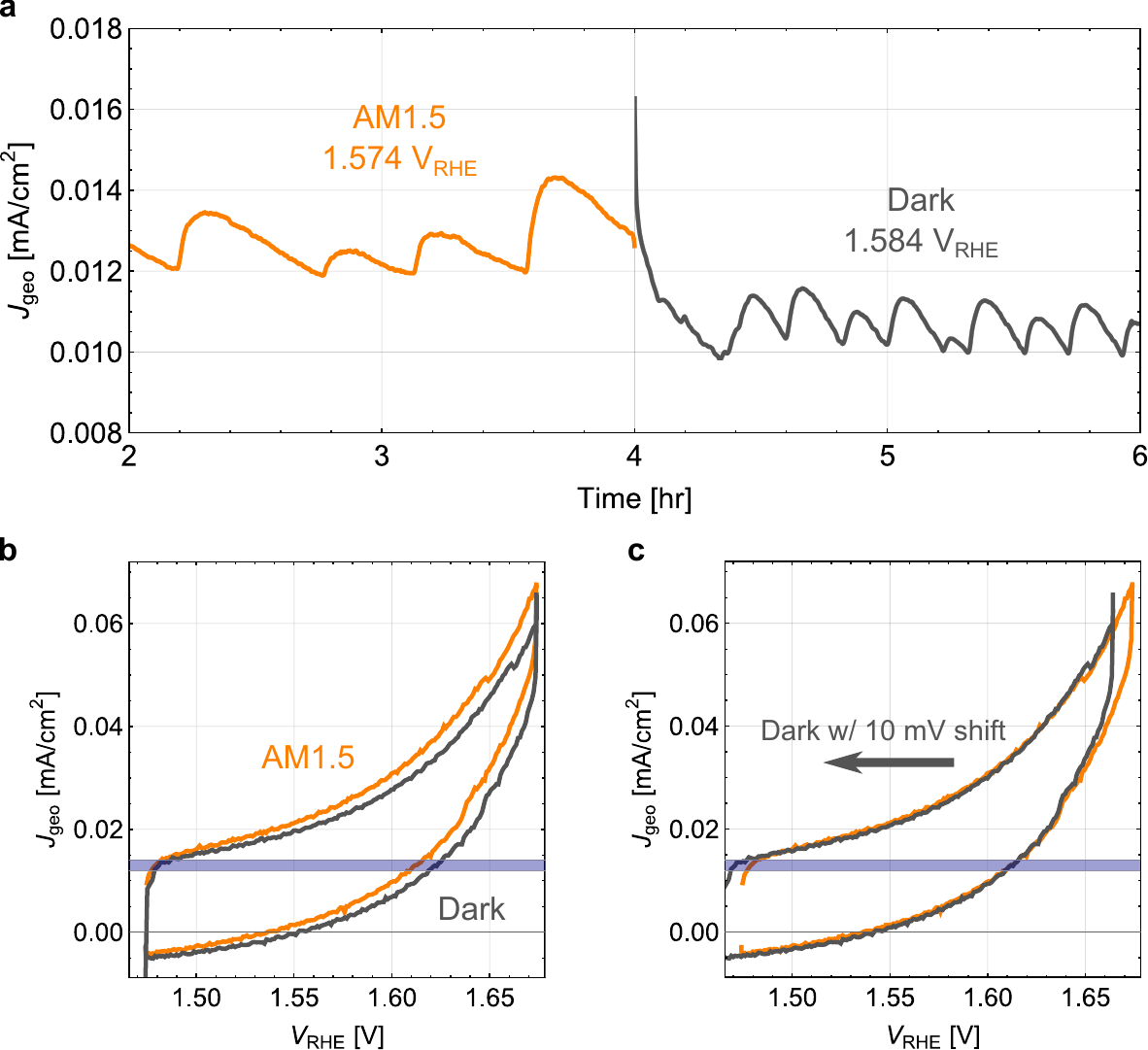}
\caption{Deactivation with Photovoltage Compensation. a) CA of a CCP electrode before and after AM1.5 illumination is removed (4 hr mark). At the time of removal, the photovoltage was increased by 10 mV, a quantity equal to the measured photovoltage from cyclic voltammetry (CV). Temperature control for this measurement was done with a hotplate with temperature probe (T = 35 $^\circ$C), and the oscillations are due to the PID loop of the hotplate. b) Stabilized CV of the same electrode as used in a), used in the determination of the appropriate photovoltage compensation. The horizontal band (blue) shows the relevant currents based on those produced in CA under AM1.5. c) The same data as b), but with a 10 mV shift to emphasize the overlap in the CV scans.
}
\end{figure}

\newpage

\begin{figure}[h!]
\centering
\includegraphics[width=0.95\textwidth]{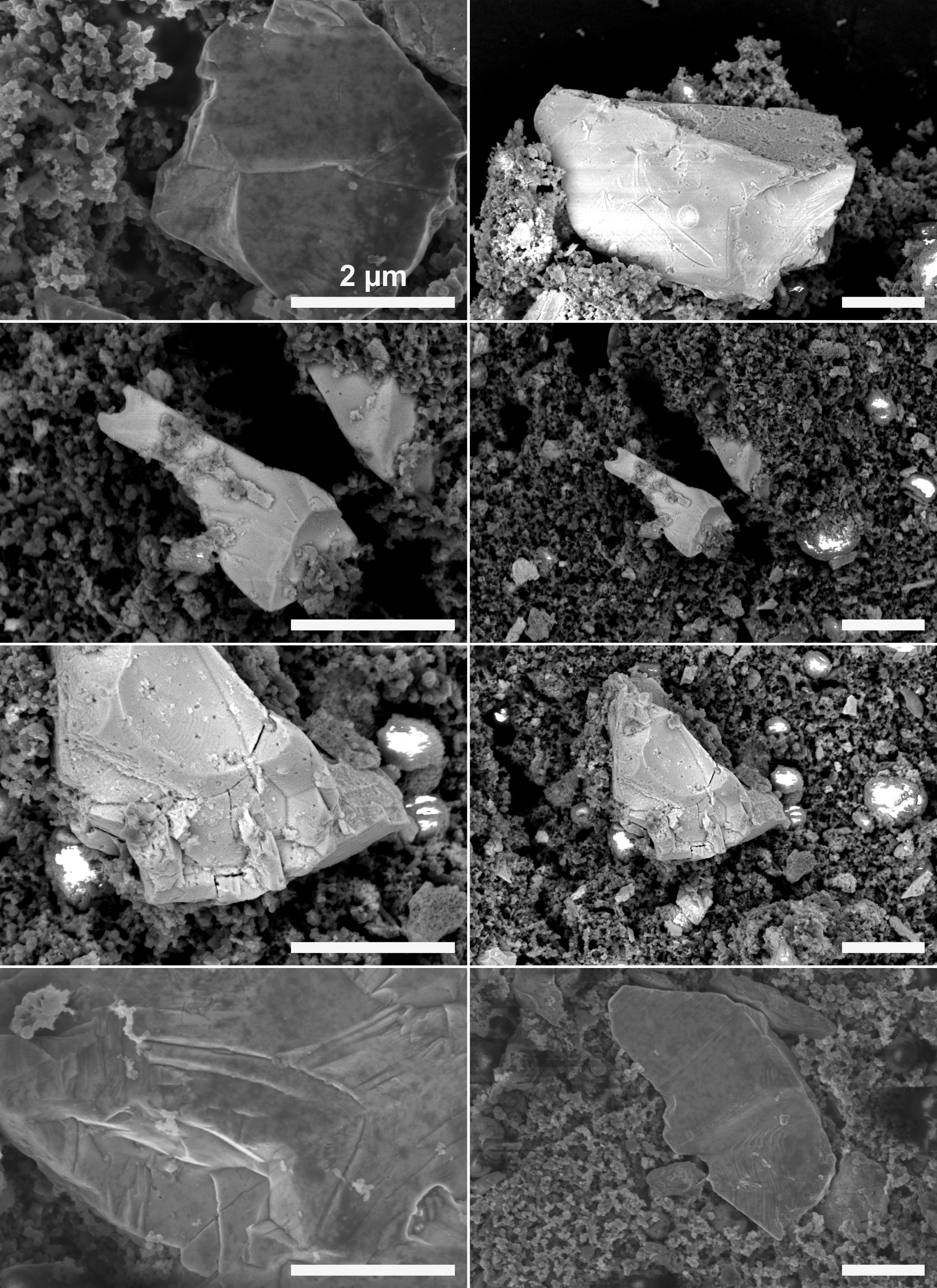}
\caption{An expanded set of SEMs of CCP particles in a freshly deposited electrode. All scale bars are 2 $\mu$m.
}
\end{figure}

\FloatBarrier

\newpage

\begin{figure}[h!]
\centering
\includegraphics[width=0.95\textwidth]{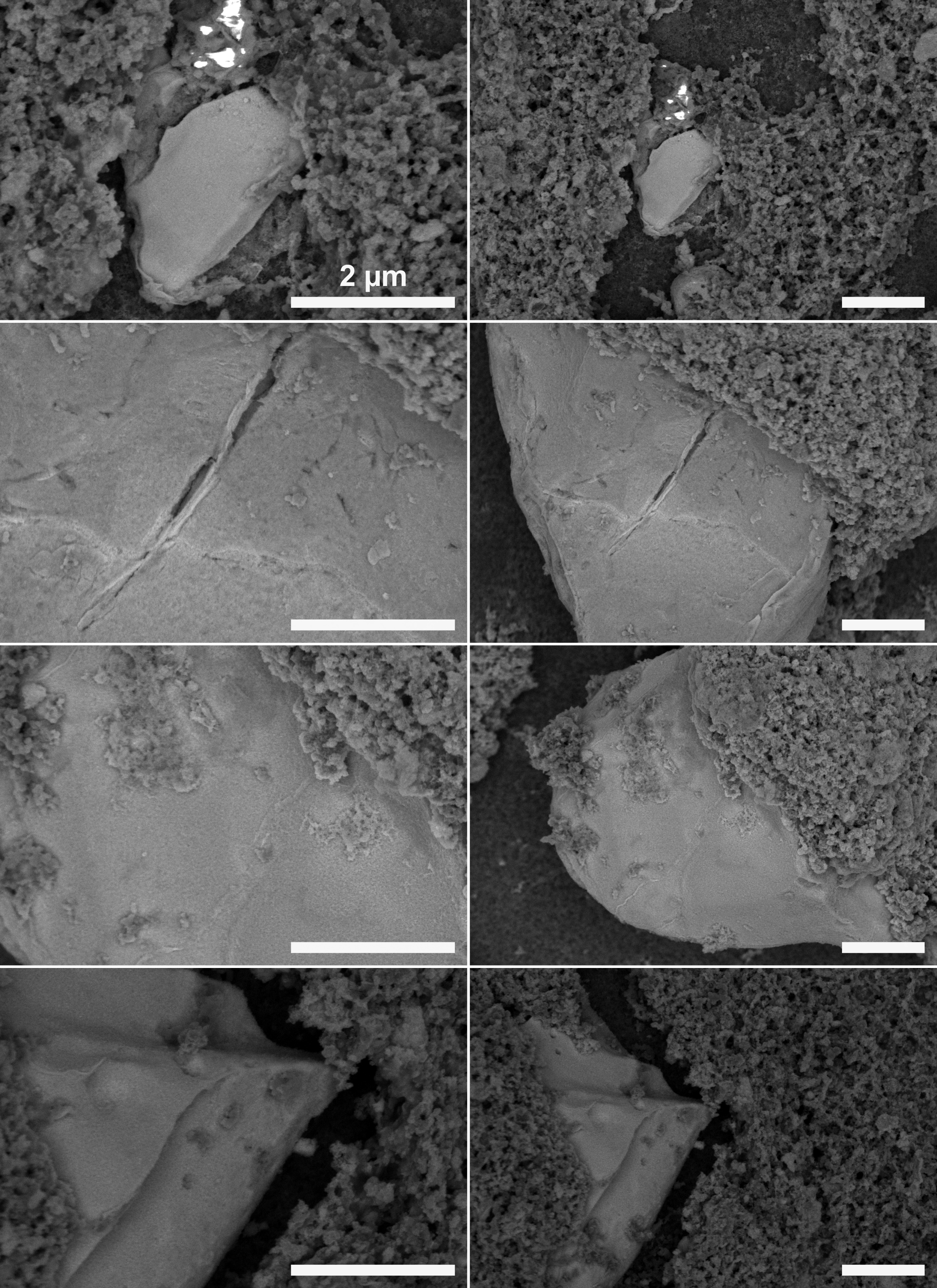}
\caption{An expanded set of SEMs of CCP particles in an electrode after 140 hr of CA at 1.6 V$_\text{RHE}$ in 0.1 M KOH under AM1.5 illumination. All scale bars are 2 $\mu$m.
}
\end{figure}

\FloatBarrier

\newpage

\begin{figure}[h!]
\centering
\includegraphics[width=0.95\textwidth]{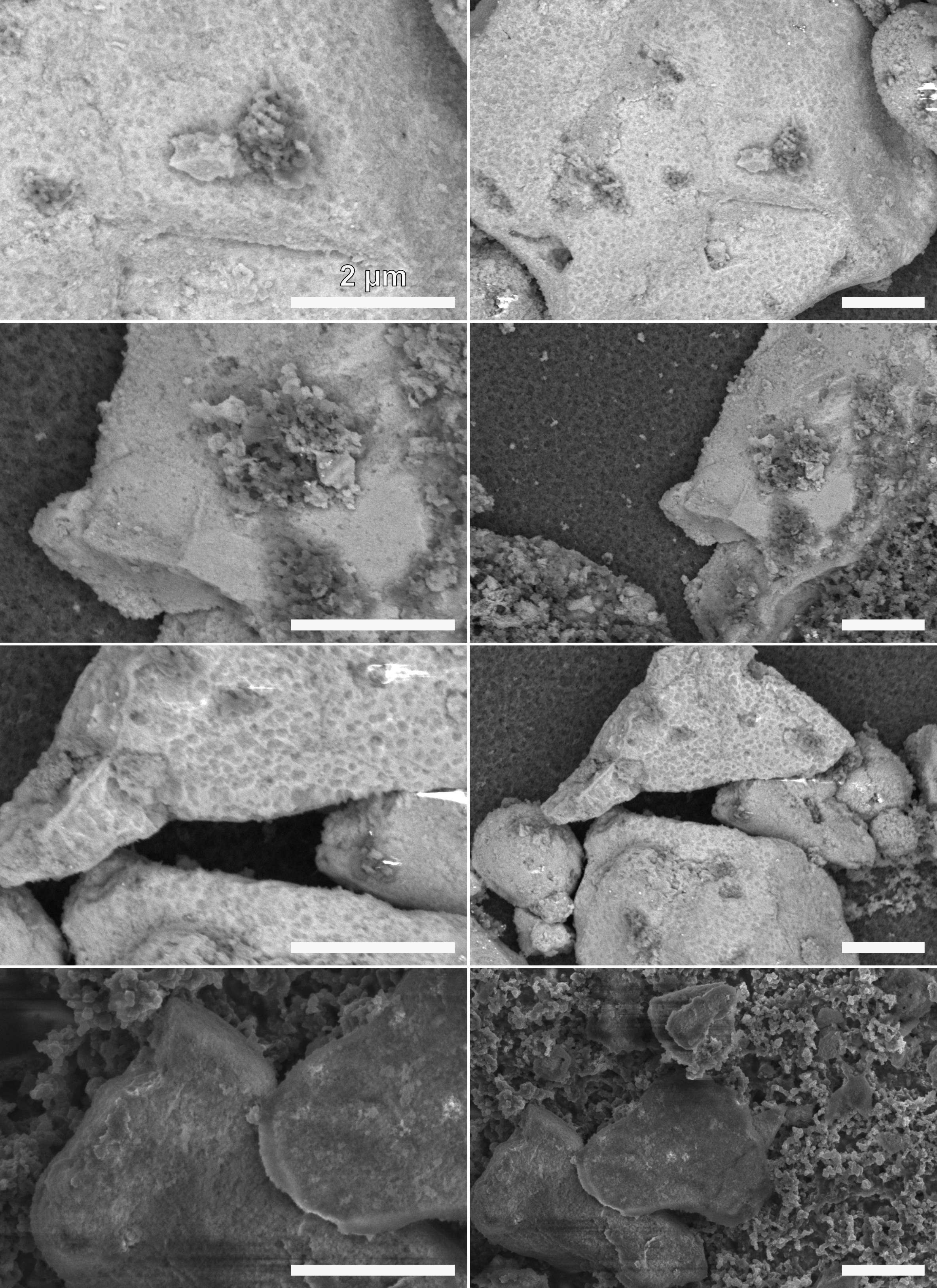}
\caption{An expanded set of SEMs of CCP particles in an electrode after 140 hr of CA at 1.6 V$_\text{RHE}$ in 0.1 M KOH in the dark. All scale bars are 2 $\mu$m.
}
\end{figure}

\FloatBarrier

\newpage

\begin{figure}[h!]
\centering
\includegraphics[width=0.65\textwidth]{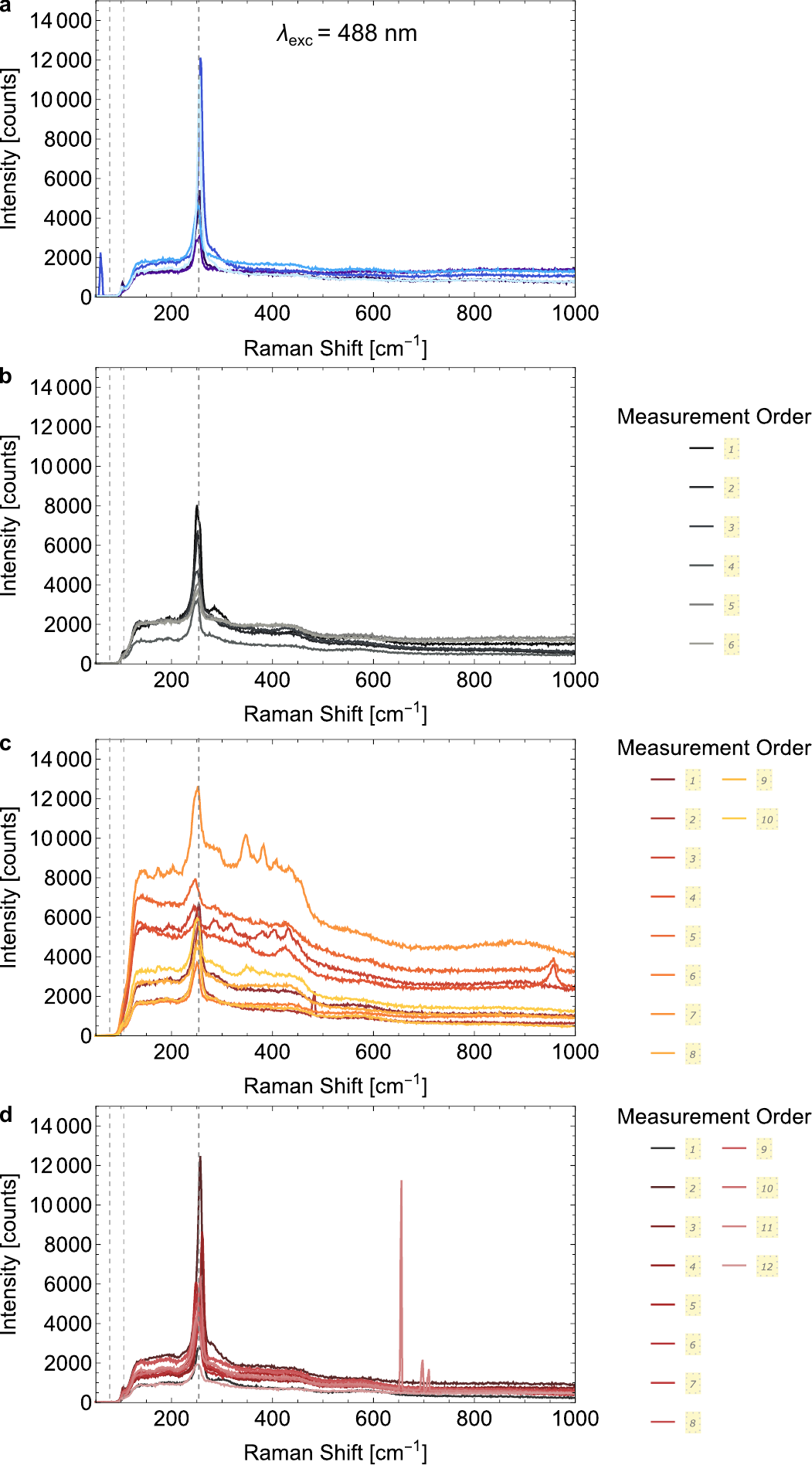}
\caption{Raman ($\lambda_\text{exc}=488$ nm) Datasets for 48 hr Chronoamperometry Experiment. a) Fresh film, b) CA in the dark, c) CA under AM1.5 illumination d) AM1.5 sample after resting in the dark for 12 hr. Each measurement within a set was performed on a different particle on the same electrode. The sequence of measurement is indicated in the legends.
}
\end{figure}

\FloatBarrier

\newpage

\begin{figure}[h!]
\centering
\includegraphics[width=0.65\textwidth]{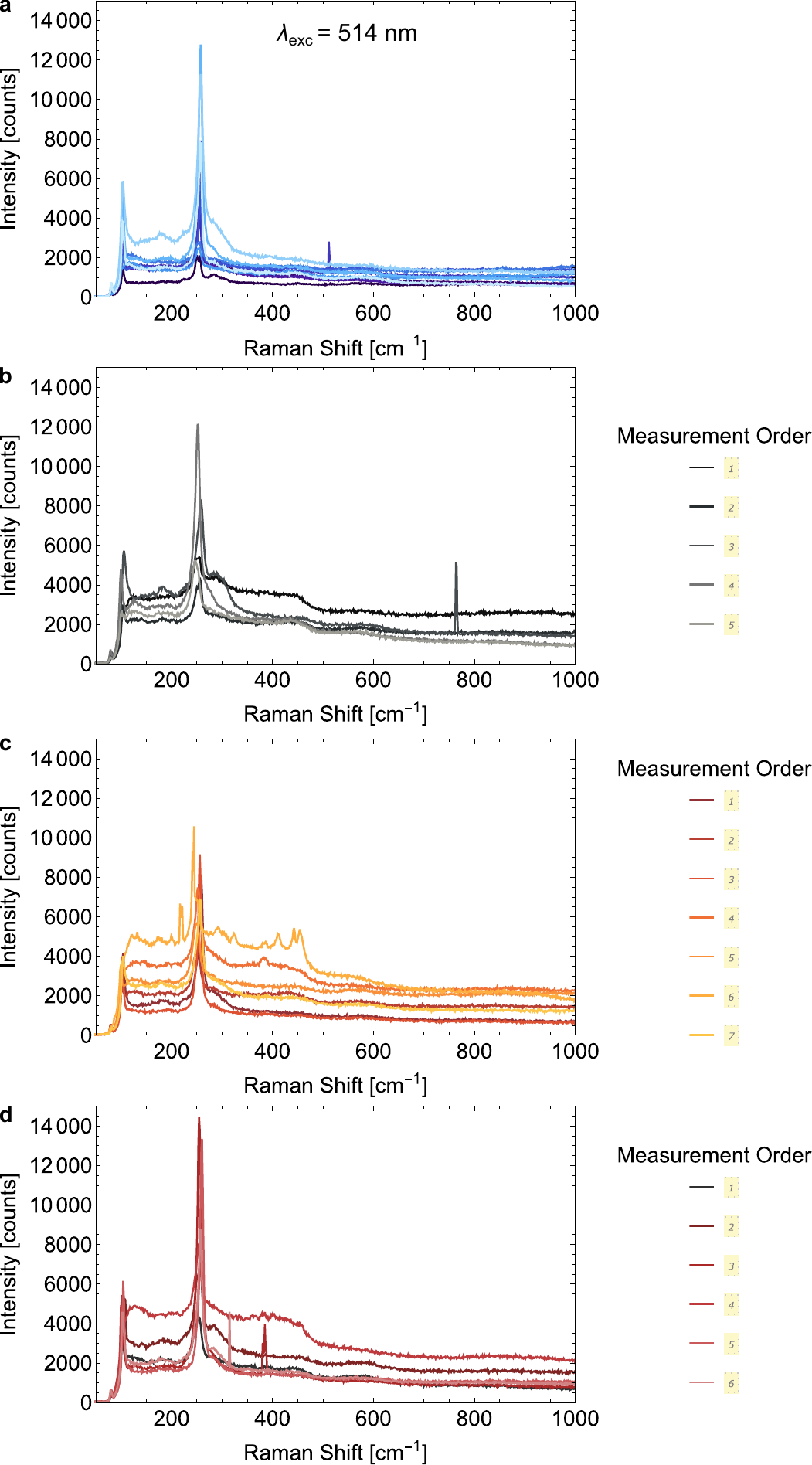}
\caption{
Raman ($\lambda_\text{exc}=514$ nm) Datasets for 48 hr Chronoamperometry Experiment. a) Fresh film, b) CA in the dark, c) CA under AM1.5 illumination d) AM1.5 sample after resting in the dark for 12 hr. Each measurement within a set was performed on a different particle on the same electrode. The sequence of measurement is indicated in the legends.
}
\end{figure}

\FloatBarrier

\newpage

\begin{figure}[h!]
\centering
\includegraphics[width=\textwidth]{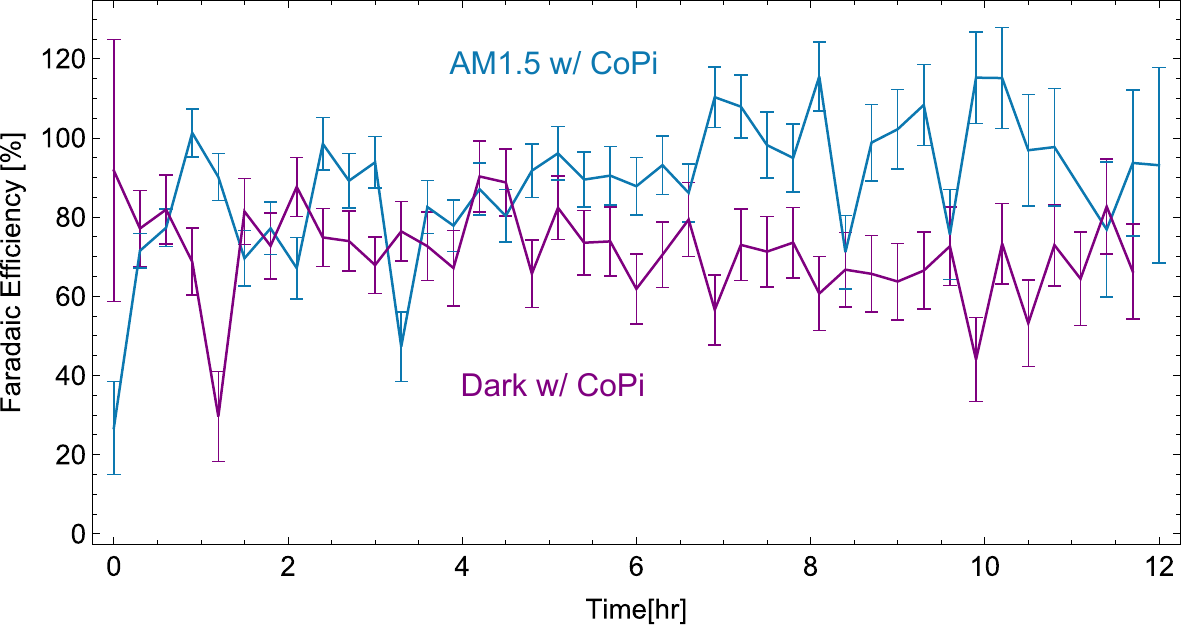}
\caption{Faradaic efficiency over time for CCP electrodes with CoPi, both collected at 1.7 V$_\text{RHE}$. Under AM1.5 illumination the activation process can be observed, in contrast to the measurement in the dark. The error bars are derived based on the uncertainty in the flow of carrier gas as well as the standard error from the calibration fit as described in the \textit{Methods}.
}
\end{figure}

\FloatBarrier

\newpage

\begin{figure}[h!]
\centering
\includegraphics[width=\textwidth]{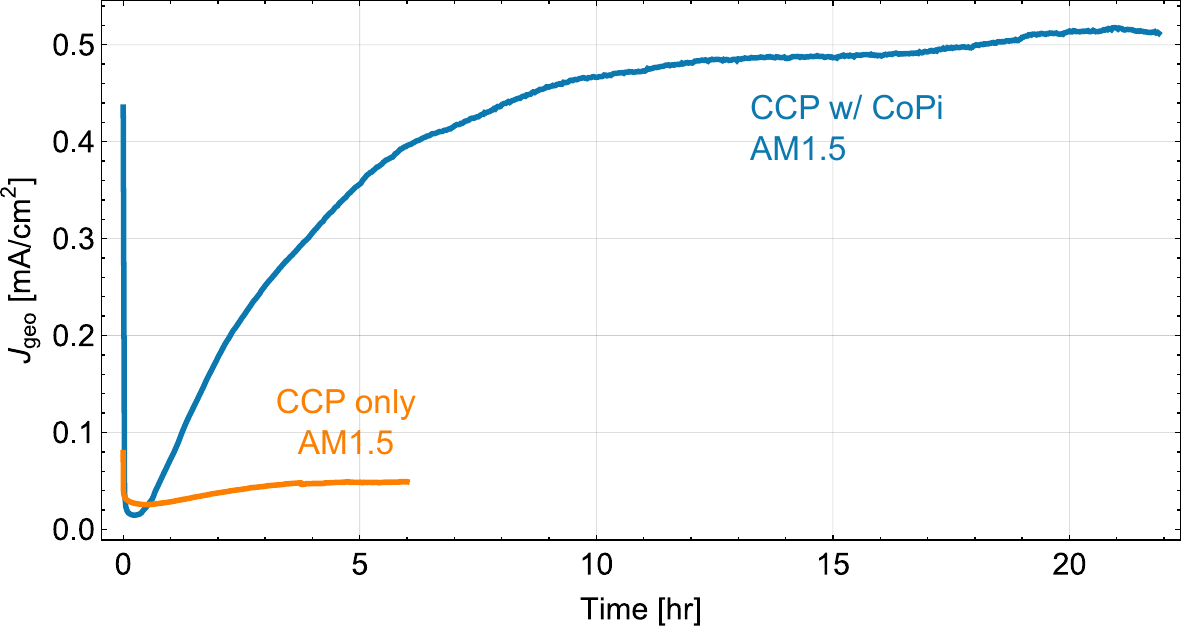}
\caption{Extended CA of CCP with CoPi electrode under AM1.5 illumination, compared against a neat CCP electrode, both collected at 1.7 V$_\text{RHE}$.
}
\end{figure}

\FloatBarrier

\newpage

\begin{figure}[h!]
\centering
\includegraphics[width=\textwidth]{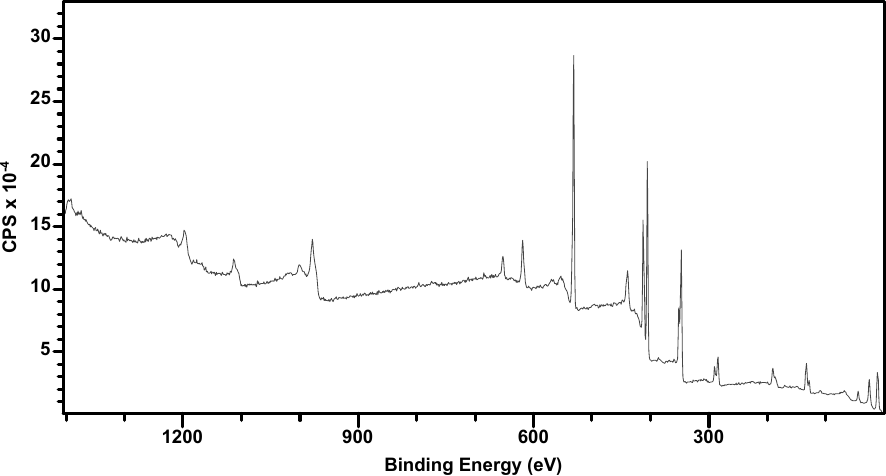}
\caption{Wide X-Ray photoelectron spectroscopy (XPS) scan of as-synthesized CCP powder.
}
\end{figure}

\newpage

\begin{figure}[h!]
\centering
\includegraphics[width=0.82\textwidth]{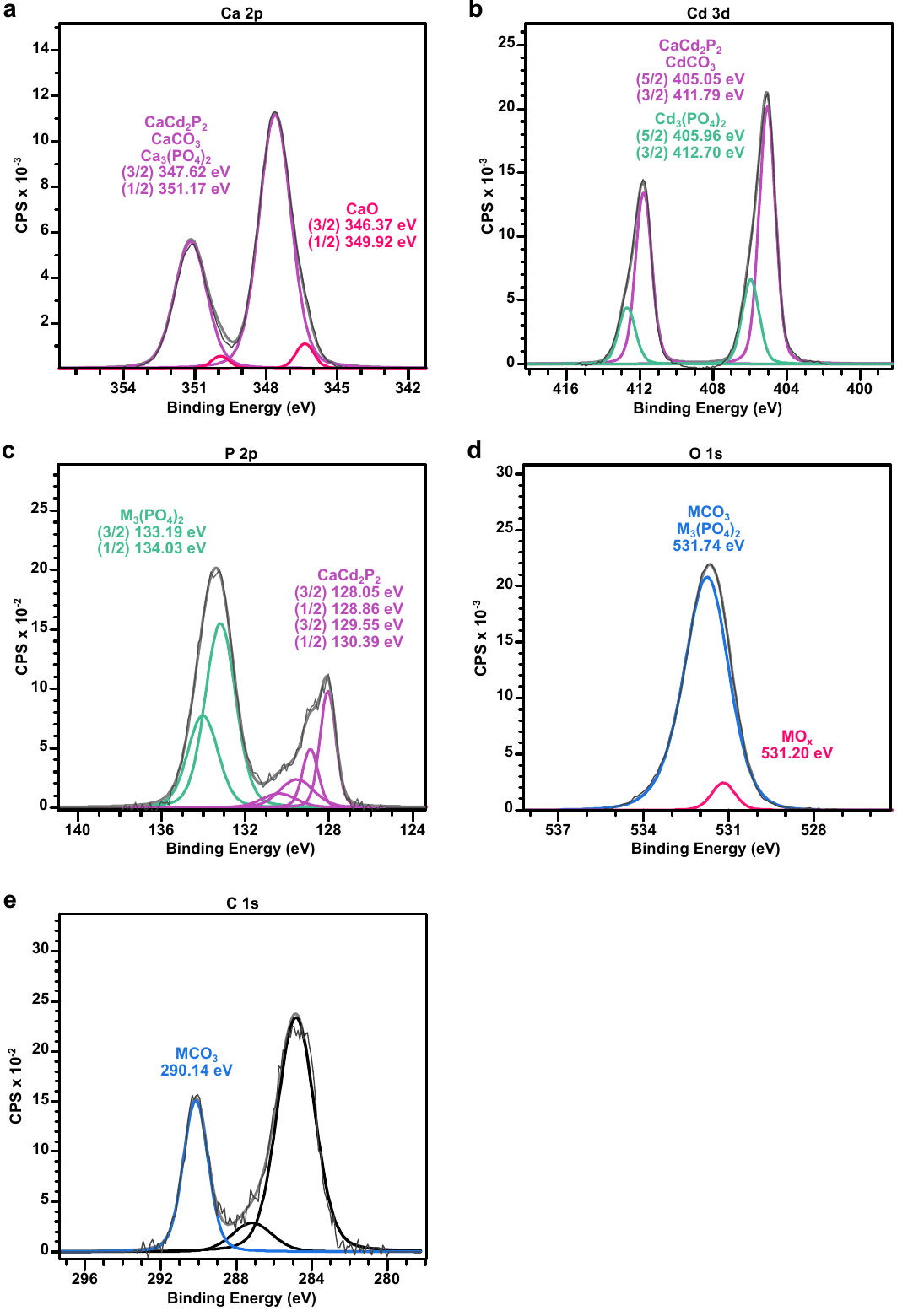}
\caption{Narrow XPS scans of as-synthesized CCP for a) Ca 2p, b) Cd 3d, c) P 2p, d) O 1s, e) C 1s. The presence of phosphates and carbonates is quite clear, while only a small proportion of signal is attributable to oxides. Raw data is in thin black line, de-convoluted peaks presented with best-fit peak centers. Doublets are presented with spin-states in parentheses. M = Ca or Cd.
}
\end{figure}

\FloatBarrier

\newpage

\begin{figure}[h!]
\centering
\includegraphics[width=\textwidth]{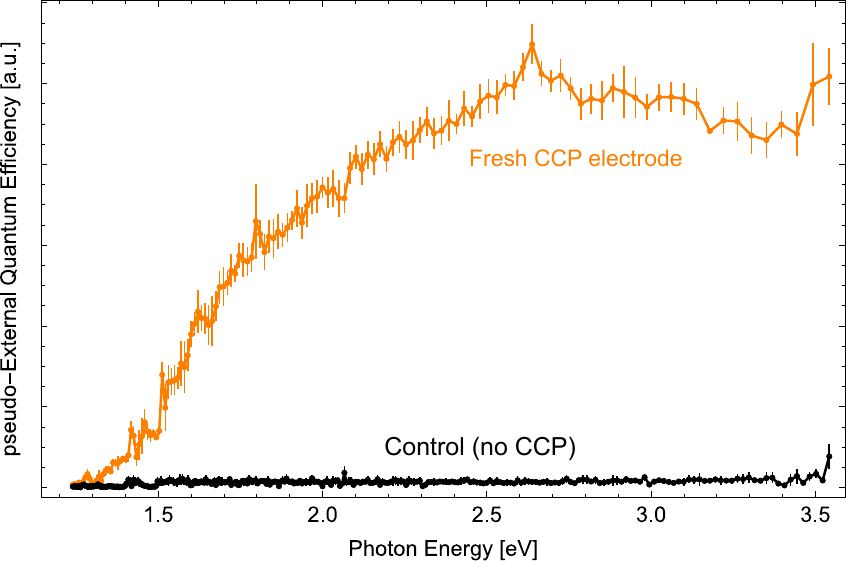}
\caption{Pseudo-External Quantum Efficiency (no area normalization, but scaled by photon count) of a fresh CCP electrode in an assembled photoelectrochemical cell under short circuit conditions, which shows the onset of current near the bandgap, with a sharp rise beyond it. This current, without bias and which increases as the wavelength and attenuation length shorten, is indicative of an activation process occurring at or near the surface. This behavior is in contrast to that of a control (no CCP) electrode. Both conditions were measured with the cell and electrode fixed at the same position and with the light visibly focused within the area of the electrode.
}
\end{figure}

\FloatBarrier

\newpage

\begin{figure}[h!]
\centering
\includegraphics[width=0.7\textwidth]{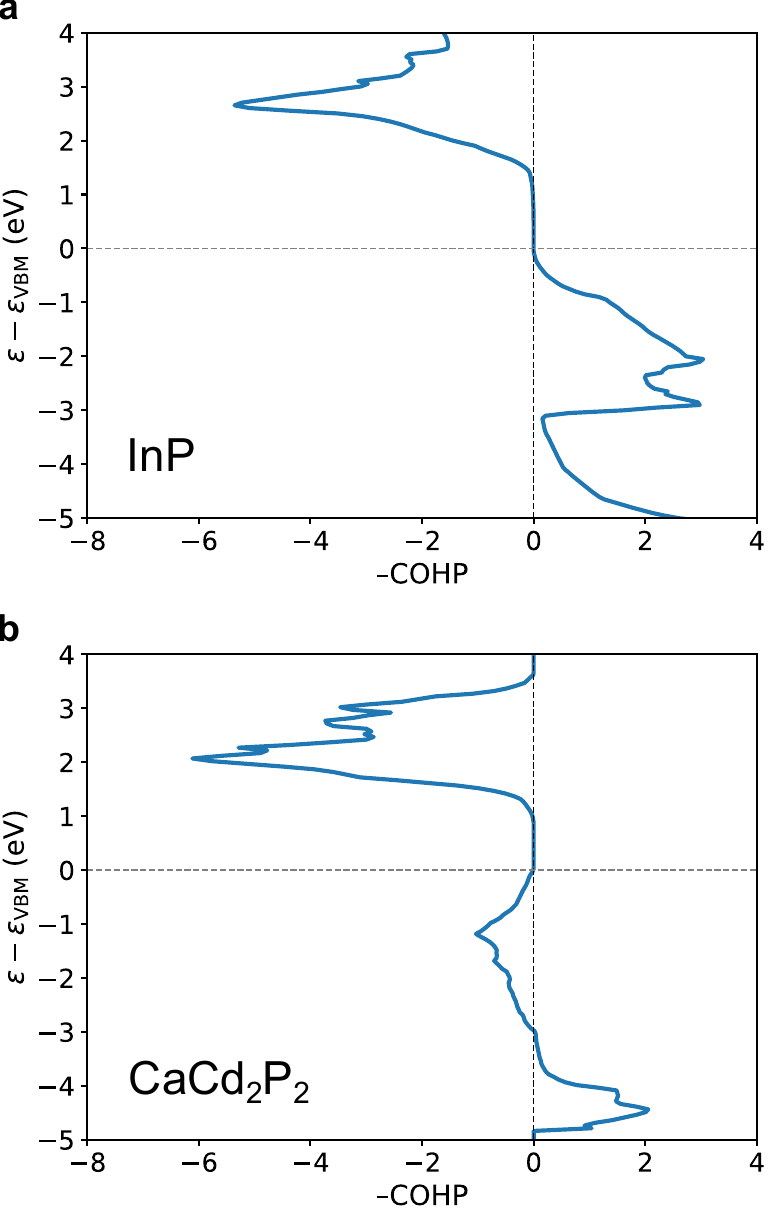}
\caption{Crystal orbital Hamilton populations (COHP) analysis. Negative and positive values of $-$COHP indicate antibonding and bonding interactions between cations and anions, respectively. a) InP is predicted to undergo a bonding to antibonding transition when an electron moves from the valence band maximum to the conduction band minimum, whereas b) CaCd$_2$P$_2$ has antibonding character at both band edges.
}
\end{figure}

\FloatBarrier

\newpage

\section{Supplementary Tables}

\newpage

\begin{table}
    \centering
    \begin{tabular}{|c|c|c|}
    \hline
    Condition &  Log$_{10} J_\text{0, geo}$  &  $J_\text{0, geo}$ [mA/cm$^2$] \\
         \hline \hline
        
        Neat, AM1.5 & -3.22 & 6.0e-4\\
     
        Neat, Dark & -3.88 & 1.3e-4\\
      
        w/ CoPi, AM1.5& -6.35 & 4.5e-7\\

        w/ CoPi, Dark& -7.06 & 8.7e-8\\
        \hline
    \end{tabular}
    \caption{Exchange Current Densities from Tafel Plot}
\end{table}

\begin{table}
    \centering
    \begin{tabular}{|c|c|c|}
    \hline
    Phase &   Mode  &  Shift [cm$^{-1}$] \\
         \hline \hline
    \multirow{4}{*}{CaCd$_2$P$_2$} & E$_\text{g}$ & 77.39\\
        
       & E$_\text{g}$ & 253.17\\
     
      &  A$_\text{1g}$ & 105.07\\
      
      &  A$_\text{1g}$ & 254.51\\
        \hline
    \multirow{6}{*}{Ca$_3$CdO$_4$} &  A$_\text{g}$ & 291.46\\
        
        & B$_\text{2g}$ & 298.46\\
        
        & B$_\text{1g}$ & 338.61\\
         
        & B$_\text{3g}$ & 338.84\\
            
        & A$_\text{g}$ & 479.32\\
        
       & B$_\text{1g}$ & 485.14\\

             \hline
      
    \multirow{4}{*}{CdH$_2$O$_2$} &   E$_\text{g}$ & 249.71\\
        
      &  E$_\text{g}$ & 796.34\\
         
       & A$_\text{1g}$ & 399.38\\
         
       & A$_\text{1g}$ & 3572.98\\
             \hline
 \multirow{1}{*}{t-CaO$_2$} & A$_\text{1g}$ & 950.646 \\
             \hline
    \end{tabular}
    \caption{Computed Raman Modes for Various Phases}
\end{table}

\FloatBarrier

\newpage

\bibliography{export}